\documentclass{article}
\usepackage[left=2cm,right=2cm,top=2cm,bottom=2cm]{geometry}

\usepackage{graphicx}

\usepackage{authblk}
\usepackage{mathptmx}      
\usepackage[utf8]{inputenc}
\usepackage[T1]{fontenc} 
\usepackage{lmodern} 

\usepackage{amsmath}
\usepackage{amsfonts}
\usepackage{amssymb}
\usepackage[labelfont=bf, labelsep=space]{caption} 
\usepackage[]{subcaption}

\usepackage{color}

\usepackage{hyperref}

\usepackage[round]{natbib}

\providecommand{\keywords}[1]
{
	\small	
	\textbf{\textit{Keywords:}} #1
}

\title{Comparing methods of modeling depth-induced breaking of irregular waves with a fully nonlinear potential flow approach }

\date{}
\author[,1]{Bruno~Simon\thanks{\href{mailto:bruno.simon@uqconnect.edu.au}{bruno.simon@uqconnect.edu.au}}}
\author[1,2]{Christos~E.~Papoutsellis\thanks{\href{mailto:cpapoutsellis@gmail.com}{cpapoutsellis@gmail.com}}}
\author[1]{Michel~Benoit\thanks{\href{mailto:benoit@irphe.univ-mrs.fr}{benoit@irphe.univ-mrs.fr}}}
\author[,2,3]{Marissa~L.~Yates\thanks{\href{mailto:marissa.yates@cerema.fr}{marissa.yates@cerema.fr}}}

\affil[1]{Aix Marseille Univ, CNRS, Centrale Marseille, Irphé  (UMR 7342), Marseille, France}
\affil[2]{Laboratoire d'Hydraulique Saint-Venant, Ecole des Ponts, Cerema, EDF, Université Paris-Est, Chatou, France}
\affil[3]{Cerema, Water, Sea and Rivers, Environment and Risks Department, Plouzané, France}

\begin{document}
		
	\maketitle
	
	\begin{abstract}
	\addcontentsline{toc}{section}{Abstract}

	The modeling of wave breaking dissipation in coastal areas is investigated with a fully nonlinear and dispersive wave model. 
	The wave propagation model is based on potential flow theory, which initially assumes non-overturning waves. 
	Including the impacts of wave breaking dissipation is however possible by implementing a wave breaking initiation criterion and dissipation mechanism. 
	Three criteria from the literature, including a geometric, kinematic, and dynamic-type criterion, are tested to determine the optimal criterion predicting the onset of wave breaking.
	Three wave breaking energy dissipation methods are also tested: the first two are based on the analogy of a breaking wave with a hydraulic jump, and the third one applies an eddy viscosity dissipative term.
	Numerical simulations are performed using combinations of the three breaking onset criteria and three dissipation methods. 
	The simulation results are compared to observations from four laboratory experiments of regular and irregular waves breaking over a submerged bar, irregular waves breaking on a beach, and irregular waves breaking over a submerged slope.
	The different breaking approaches provide similar results after proper calibration.
	The wave transformation observed in the experiments is reproduced well, with better results for the case of regular waves than irregular waves.
	Moreover, the wave statistics and wave spectra are predicted well in general, and in particular for regular waves. 
	Some differences are observed for irregular wave cases, in particular in the low-frequency range. 
	This is attributed to incomplete absorption of the long waves in the numerical model. 
	Otherwise, the wave spectra in the range $[0.5f_p,\: 5f_p]$ are reproduced well, before, inside, and after the breaking zone for the three irregular wave experiments.
		
	\end{abstract}
	

	\section{Introduction}
	
	As ocean waves propagate toward the shore, they evolve under the effects of a variety of physical processes, among which shoaling, refraction and breaking have dominant roles.
	Accurate modeling of these processes requires a mathematical approach that properly reproduces both dispersive and nonlinear effects.
	One option is to use Computational Fluid Dynamics (CFD) codes solving the Navier-Stokes equations to simulate the physical processes in detail, but this approach is computationally expensive and can be applied at local scales only.
	
	Another faster method consists in solving approximate equations, such as one form of the Boussinesq or Serre-Green-Naghdi equations.
	However, these mathematical systems are unable to retain the complete dispersive and nonlinear properties of the water wave problem.
	
	Extension of the Nonlinear Schrödinger equations \citep{dysthe_note_1979, kharif_physical_2003} can be used to capture nonlinearity and dispersion for narrow-banded sea-states.
	A more general approach consists in solving a fully nonlinear and dispersive potential flow model, which can be reformulated as the Zakharov equations \citep{zakharov_stability_1968, craig_numerical_1993}.
	This approach resolves the potential flow problem by advancing in time free surface quantities only.
	Several approaches can be considered to solve the Zakharov equations, such as the Hamiltonian Coupled-Mode Theory (or HCMT) method \citep{athanassoulis_exact_2017, papoutsellis_implementation_2018}, the extension of the high-order spectral (or HOS) method to variable bottom cases \citep{gouin_development_2016}, or the direct use of finite difference schemes \citep{bingham_accuracy_2007}.
	Other commonly applied approaches for solving the fully nonlinear potential flow problems are, for example, the Boundary Element Method \citep{longuet-higgins_deformation_1976, dold_efficient_1986, grilli_efficient_1989, dold_efficient_1992}, the Finite Element Method \citep{wu_finite_1994}, the Quasi-Arbitrary Lagrangian Eulerian Finite Element Method \citep{ma_quasi_2006}, the Spectral Element Method \citep{engsig-karup_stabilised_2016} and the Spectral Boundary Integral Method \citep{fructus_efficient_2005, wang_numerical_2015}.
	
	In this study, a spectral approach is applied in the vertical direction using Chebyshev polynomials, following \citet{tian_numerical_2008}.
	This approach is accurate and converges quickly as a function of the number of polynomials used, as previously demonstrated \citep{yates_accuracy_2015, raoult_validation_2016, benoit_analysis_2017}.
	One limitation of the Zakharov equations is the assumption of a non-overturning free surface, which precludes direct modeling of wave breaking.
	In order to overcome this limitation, some of the effects of wave breaking can be simulated by incorporating a term in the dynamic and/or kinematic free surface boundary conditions dissipating energy during wave breaking.
	
	Several approaches have been developed to model wave energy dissipation after a criterion detects the onset of breaking. 
	Among the different criteria used, the most common approaches are based on considering: the angle of the wave front \citep{schaffer_boussinesq_1993}, the vertical velocity at the free surface \citep{kennedy_boussinesq_2000, kurnia_high_2014}, a combination of both \citep{kazolea_numerical_2014}, or the relative trough Froude number \citep{okamoto_relative_2006}. 
	Wave breaking can also be detected by computing the ratio of the horizontal wave crest particle velocity over wave phase speed, also defined as the energy flux velocity at the crest divided by crest velocity by \citet{barthelemy_unified_2018}.
	
	Following the detection of the onset of wave breaking, several different dissipation mechanisms can be applied by adding a  term to the dynamic free surface boundary condition.
	Over the specified spatial extent of wave breaking, the wave energy dissipation can be calculated using a surface roller model \citep{schaffer_boussinesq_1993}, a vorticity model \citep{svendsen_wave_1978}, an eddy viscosity model \citep{kennedy_boussinesq_2000, kurnia_high_2014, papoutsellis_fully_2018, papoutsellis_modeling_2019}, or by analogy with a hydraulic jump \citep{guignard_modeling_2001, grilli_fully_2019, papoutsellis_modeling_2019}.
	These mechanisms require parameterizing both the definition of the onset of wave breaking and the intensity and spatial extent of the applied dissipation.
	
	In this work, three wave breaking criteria and three dissipation mechanisms are implemented in a numerical code solving the Zakharov equations (code whispers3D).
	This work builds on previous work using the wave model presented in \citet{simon_modeling_2018} which has been extended with additional breaking criteria and dissipation mechanisms, as well as additional irregular wave breaking cases.
	It also builds on a similar fully nonlinear and dispersive potential model \citep{papoutsellis_modeling_2019} solving the same mathematical model (Zakharov equations), but using the HCMT approach. 
	The focus of the current study is to evaluate the accuracy of the breaking criteria and wave energy dissipation mechanisms in reproducing the propagation and transformation of irregular waves over various types of coastal bathymetries.
	The simulation results are compared to four sets of experimental measurements confirming the applicability of the proposed modeling strategies, both for regular and irregular wave conditions.

	\section{Mathematical modeling and numerical methods}
		\subsection{Overview of the system of equations}
		For an inviscid and homogeneous fluid of constant density, potential flow theory can be used if the flow is assumed irrotational.
		In the following, the domain is restricted to two dimensions in the vertical plane $(x,z)$, with $x$  the horizontal axis and $z$ the vertical axis (positive upward), with $z=0$ at the still water surface.
		In this case, the velocity vector $\underline u(x,z,t) = (u,w)$ can be expressed as the gradient of the velocity potential $\Phi(x,z,t)$: $\underline u=\nabla\Phi=(\partial_x\Phi,\partial_z\Phi)$.
		The velocity potential must then satisfy the Laplace equation in the fluid domain: $\Delta \Phi= \Phi_{xx}+ \Phi_{zz}= 0$.
		\\
		The Laplace equation is supplemented with boundary conditions at the free surface, bottom, and lateral boundaries.
		The seabed, $z=-h(x)$, is considered impermeable, fixed and smooth, such that:
		\begin{equation}
			\partial_x\Phi \partial_x h + \partial_z \Phi=0, \quad \text{on }\ z=-h(x).
		\end{equation}
		The free surface, $z=\eta(x,t)$, is continuous and  assumed non-overturning.
		At the lateral boundaries, Dirichlet or Neumann boundary conditions are defined.
		After assuming uniform atmospheric pressure at the free surface (by convention set to $0$), and defining the velocity potential at the free surface as $\psi(x,t):=\Phi(x, z=\eta(x,t),t)$, the nonlinear free surface boundary conditions are formulated as the so-called Zakharov equations \citep{zakharov_stability_1968, craig_numerical_1993}:
		\begin{align}
			\partial_t \eta &=- \partial_x \eta \partial_x\psi + W (1+( \partial_x \eta)^2),  \label{eq:zak1} \\
			\partial_t \psi &=-g\eta-\frac{1}{2}(\partial_x\psi)^2 + \frac{1}{2} W^2(1+(\partial_x\eta)^2), \label{eq:zak2}
		\end{align}
		where $W(x,t)$ is the vertical velocity at the free surface $W(x,t)=\partial_z \Phi(x, z=\eta(x,t), t)$, and $g$ is the acceleration of gravity.
		
		Eqs. \eqref{eq:zak1}-\eqref{eq:zak2} only involve  free surface variables, depending on $x$ and $t$.
		To integrate these equations in time, a Dirichlet-to-Neumann (DtN) problem must be solved to determine $W$ from $(\eta,\psi)$, as described in \citet{yates_accuracy_2015}, \citet{raoult_validation_2016} and \citet{benoit_analysis_2017} and summarized in the next subsection.
		
		\subsection{Numerical solution of the DtN problem with a spectral method}
		
		The DtN problem to be solved, at a given time $t$, is a Laplace boundary value problem (BVP) for the potential $\Phi$ in the fluid domain, composed of:
		\begin{align}
			\Phi_{xx} + \Phi_{zz} & =0 \quad       &-h(x)\le z \le \eta(x,t), \label{eq:BVP1}  \\
			\Phi(x,\eta) &          =\psi(x) \quad &\text{on }z=\eta(x,t), \label{eq:BVP2} \\
			h_x\Phi_x + \Phi_z &     =0 \quad      &\text{on } z=-h(x), \label{eq:BVP3}
		\end{align}
		completed by lateral boundary conditions.
		A numerical code, called whispers3D, is being developed at Centrale Marseille and the Irphé laboratory to solve the above BVP problem Eqs. \eqref{eq:BVP1}-\eqref{eq:BVP3} and to march Eqs. \eqref{eq:zak1}-\eqref{eq:zak2} in time. 
		It combines high-order finite difference schemes to approximate the horizontal derivatives with a spectral method in the vertical direction using the Chebyshev-tau approach, following the work of \citet{tian_numerical_2008} and \citet{yates_accuracy_2015}. 
		
		With this method, the velocity potential is approximated using a linear combination of Chebyshev polynomials of the first kind, $T_n(s)$, $n=0,1,..., N_T$ for $s \in [-1, 1]$:
		\begin{align}
			\Phi(x,z) = \varphi(x,s)\approx \sum_{n=0}^{N_T}a_n(x) T_n(s), \label{potential}
		\end{align}
		where $s$ is  a scaled vertical coordinate, varying from $-1$ at the bottom to $+1$ at the free surface:
		\begin{align}
			s(x,z,t)=\frac{2z+h(x)-\eta(x,t)}{h(x)+\eta(x,t)}, \label{stransform}
		\end{align}
		and $N_T$ is the maximum order of the Chebyshev polynomials used in the approximation. 
		In the following simulations, $N_T$ is set to $7$, which was found to be a good compromise between computational speed and accuracy for non breaking cases \citep{yates_accuracy_2015, raoult_validation_2016} and which was verified also for breaking cases.
		
		The BVP problem \eqref{eq:BVP1}-\eqref{eq:BVP3} is first reformulated in the $(x,s)$ coordinate system, then the approximation of the velocity potential, Eq. \eqref{potential}, is inserted in these equations. 
		After applying a Galerkin-like approach in the vertical direction to the Laplace Eq.~\eqref{eq:BVP1} and using the orthogonality properties of the Chebyshev polynomials, in combination with the two boundary conditions (Eq.~\eqref{eq:BVP2} at the free surface, $s=+1$, and  Eq.~\eqref{eq:BVP3} at the bottom, $s=-1$), a linear system of $N_T+1$ equations with unknowns $a_n(x)$, $n=0,1,..., N_T$ is formed. 
		In the current version of whispers3D, this system is solved using a GMRES algorithm with Incomplete L-U preconditioning.
		Once the $a_n$ coefficients are known, the vertical velocity at the free surface is readily obtained as:
		\begin{align}
			W(x,t) = \frac{2}{h(x)+\eta(x,t)} \sum_{n=1}^{N_T} a_n(x,t) n^2, \label{vitesse_vert}
		\end{align}
		and Eqs. \eqref{eq:zak1}-\eqref{eq:zak2} are integrated in time using an explicit third-order strong stability preserving Runge-Kutta scheme (SSP-RK($3,3$))  with constant time step \citep{gottlieb_high_2005}.
		
		At the left boundary, incident waves are generated by reconstructing $\eta$ and $\psi$ from a measured wave signal (for instance, a wave gauge from experiments) by decomposing it as a sum of independent sinusoidal waves using linear wave theory. 
		A relaxation zone adjacent to the lateral boundary progressively imposes the incident wave conditions in the domain, while absorbing reflected waves propagating toward the boundary. 
		At the opposite end of the domain, an absorbing relaxation zone forces the solution of $\eta$ and $\psi$ to $0$ with a progressive transition in space to dampen waves and limit wave reflections.
				
		\subsection{Modeling wave breaking}

		Pursuing preliminary developments made in whispers3D \citep{simon_modeling_2018}, wave breaking is modeled with a combination of two steps. 
		In the first step, a breaking criterion detects the onset (and possibly termination) of wave breaking.
		In the second step, the wave energy dissipation is simulated using a mechanism that computes a dissipative term $D$ that is added to the dynamic free surface boundary condition, Eq. \eqref{eq:zak2}.
		
		\subsubsection{Wave breaking detection criteria}
\begin{figure}[h!t]
			\centering
			\includegraphics[width=84mm]{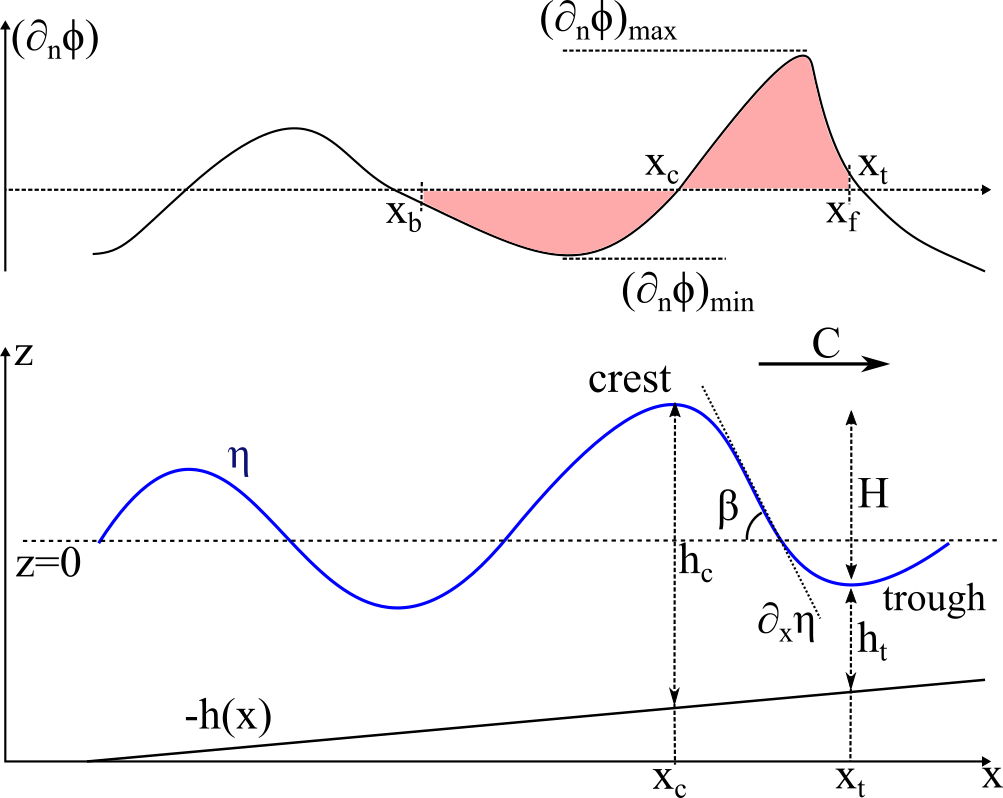}
			\caption{Definition sketch of geometric and kinematic parameters used to detect breaking waves, adapted from \cite{guignard_modeling_2001}}
			\label{fig:sketchwave}
		\end{figure}

		In this work, three of the most studied wave breaking onset criteria are tested. 
		The first type of criteria is geometric and uses the slope of the wave front to initiate wave breaking \citep{schaffer_boussinesq_1993}. 
		Wave breaking is activated when the angle of the wave slope with the horizontal, denoted $\beta$ (see \figurename{} \ref{fig:sketchwave}), exceeds a threshold value: $\beta_{i}$. 
		In the literature, the value of $\beta_{i}$ ranges from $14^\circ$ to $38 ^\circ$, depending on the type of breaking (e.g. plunging, spilling) and the local bottom slope \citep{okamoto_relative_2006}. 
		Using a Boussinesq-type model, \citet{cienfuegos_wave-breaking_2010} showed that the calibration of this criterion depends on the wave field, and they recommended  using $\beta_{i}$ between $28^\circ$ and $32^\circ$ for spilling breakers, and between $35^\circ$ and $36^\circ$ for plunging breakers.
		The slope angle can also be used to stop the dissipation when wave breaking terminates. 
		The angle of termination, denoted $\beta_{t}$, is usually set to approximately $10^\circ$ \citep{schaffer_boussinesq_1993, cienfuegos_wave-breaking_2010}.
		\\
		
		The second type of criteria is kinematic. 
		The wave breaking onset criterion uses the ratio between the normal velocity $\partial_t\eta$ on the wave front over the corresponding wave speed in shallow water $\sqrt{gh}$.
		Breaking is initiated when $\partial_t\eta/\sqrt{gh}>\gamma_i$ and is terminated when $\partial_t\eta/\sqrt{gh}<\gamma_t$ for a breaking wave.
		The parameters $\gamma_i$ and $\gamma_t$  are calibrated with the experimental data \citep{okamoto_relative_2006, kurnia_high_2014}, although authors using these parameters do not mention how to select values depending on the wave field or type of breaking.
		The most commonly used thresholds to detect breaking are $\gamma_i\in [0.35, 0.65]$ \citep{kennedy_boussinesq_2000, kazolea_numerical_2014}.
		Termination threshold values are usually around $\gamma_t \approx 0.15$ \citep{kennedy_boussinesq_2000, kurnia_high_2014}.
		A series of  tests showed little sensitivity of the results to parameters varying in the recommended interval.
		\\
		
		The third type of criteria is dynamic and is based on the local energy flux velocity recently presented in \citet{barthelemy_unified_2018}. 
		For two dimensional flows, this breaking criterion can be reduced to a dynamical criterion: $B_x=u_1(x_c,t)/C$, where $C$ is the (local) crest velocity, $x_c$ the position of the crest, and $u_1$ the horizontal orbital velocity at the crest. 
		Wave breaking is detected when $B_x=u_1(x_c,t)/C > \sigma_{i}$. 
		\citet{kurnia_high_2014} proposed $\sigma_{i} \in [0.7, 1]$, whereas \citet{barthelemy_unified_2018} suggest that breaking occurs when $B_x$ is larger than $[0.85, 0.86]$ for waves in deep and intermediate water depths.
		The same authors expect a similar range of values to be valid for shallow water conditions.
		\citet{saket_threshold_2017} re-examined this threshold experimentally using thermal image velocimetry, and they reduced the proposed range of $\sigma_{i}$ to $0.84\pm 0.016$.
		However, recent work concluded that the wave breaking onset should be $B_x > 0.90$ for intermediate water wave groups specifically generated to produce fewer spurious free waves \citep{hasan_evaluation_2019}.
		\\
		
		\figurename{} \ref{fig:sketchwave} presents the characteristics of a breaking wave with celerity $C$.
		$h_c$ is the water depth under the crest located at $x_c$, $h_t$ is the depth under the front trough at $x_t$, $H$ is the wave height, and $\beta$ is the angle of the slope of the wave front relative to the horizontal. 
		The interval $[x_b, x_f]$ is the zone of energy dissipation, as defined by \citet{guignard_modeling_2001}.
		
		\subsubsection{Wave energy dissipation mechanisms}
		Once the onset of wave breaking is detected, a dissipation term is applied over a defined spatial extent of the breaking wave. 
		Three dissipation mechanisms are considered here, all relying on adding a dissipation term, denoted $D$, to the right-hand-side of Eq. \eqref{eq:zak2}.
		\\
		The first method (denoted HJ in the following) uses an analogy with a hydraulic jump propagating over a flat bottom to estimate a pressure term $D = -P_{surf} / \rho$, with $\rho$ the water density \citep{guignard_modeling_2001, grilli_fully_2019}. 
		The zone where the pressure is applied extends on both sides of the breaking wave crest between $x_f$ and $x_b$ (see \figurename{} \ref{fig:sketchwave}) determined by:
		\begin{equation}
			\left| \partial_n\Phi / \left(\partial_n\Phi\right)_{\min} \right| >\epsilon \text{ and } 	\left| \partial_n\Phi / \left(\partial_n\Phi\right)_{\max} \right| >\epsilon \quad \text{for} \quad x \in \left[x_b, x_f\right], 
		\end{equation}
		with $(\partial_n \Phi)_{\min}$ and $(\partial_n \Phi)_{\max}$ the minimum (respectively maximum) value of $\partial_n \Phi$ in the back (respectively front) of the wave (\figurename{}~\ref{fig:sketchwave}) and $\epsilon$ a small threshold value (for instance, $\epsilon=10^{-5}$ in \citet{guignard_modeling_2001}). 
		The normal derivative of the potential $\Phi$ at the free surface is:
		\begin{equation}
			\partial_n \Phi (\eta) = \underline n \cdot \nabla\Phi = \partial_t \eta = -\partial_x \eta \partial_x \psi +W(1+(\partial_x \eta)^2),
		\end{equation}
		where $\underline n$ is the normal vector (non-unit) to the free surface.
		\\
		The pressure $P_{surf}$, non-zero in the breaker zone $[x_f, x_b]$, is computed as:
		\begin{equation}
			P_{surf}=\nu_b S(x) \partial_t\eta(x),
		\end{equation}
		where $S(x)$ is a function ensuring a smooth spatial transition in areas where breaking dissipation is applied. 
		It varies sinusoidally from $0$ to $1$ at the defined limits $[x_b,x_{b1}]$ and $[x_{f1},x_f]$ of the breaking wave, such that $x_{b1}=x_b-\alpha(x_b - x_f)$ and $x_{f1}=x_f - \alpha(x_b - x_f)$ ($\alpha$ is defined depending on the case, and $\alpha=0.1$ is used here, as suggested by \citet{guignard_modeling_2001}).
		The absorption coefficient, $\nu_b$, is defined as:
		\begin{align}
			\nu_b=\mu \Pi_{jump} \left(\int_{x_b}^{x_f}{S(x) \left(\partial_t\eta(x)\right)^2 dx} \right)^{-1},
		\end{align}
		where $\mu$ is a coefficient controlling the magnitude of the dissipation.
		\citet{svendsen_wave_1978} recommend selecting values close to $1.5$ for $\mu$. 
		$\Pi_{jump}$ is the power dissipated in a turbulent bore \citep{lamb_hydrodynamics_1932}, adapted by \citet{guignard_modeling_2001} as:
		\begin{equation}
			\Pi_{jump}= \rho g C \dfrac{h H^3}{4 h_c h_t}.
		\end{equation}
		
\begin{figure}[h!t]
			\centering
			\includegraphics[width=\linewidth]{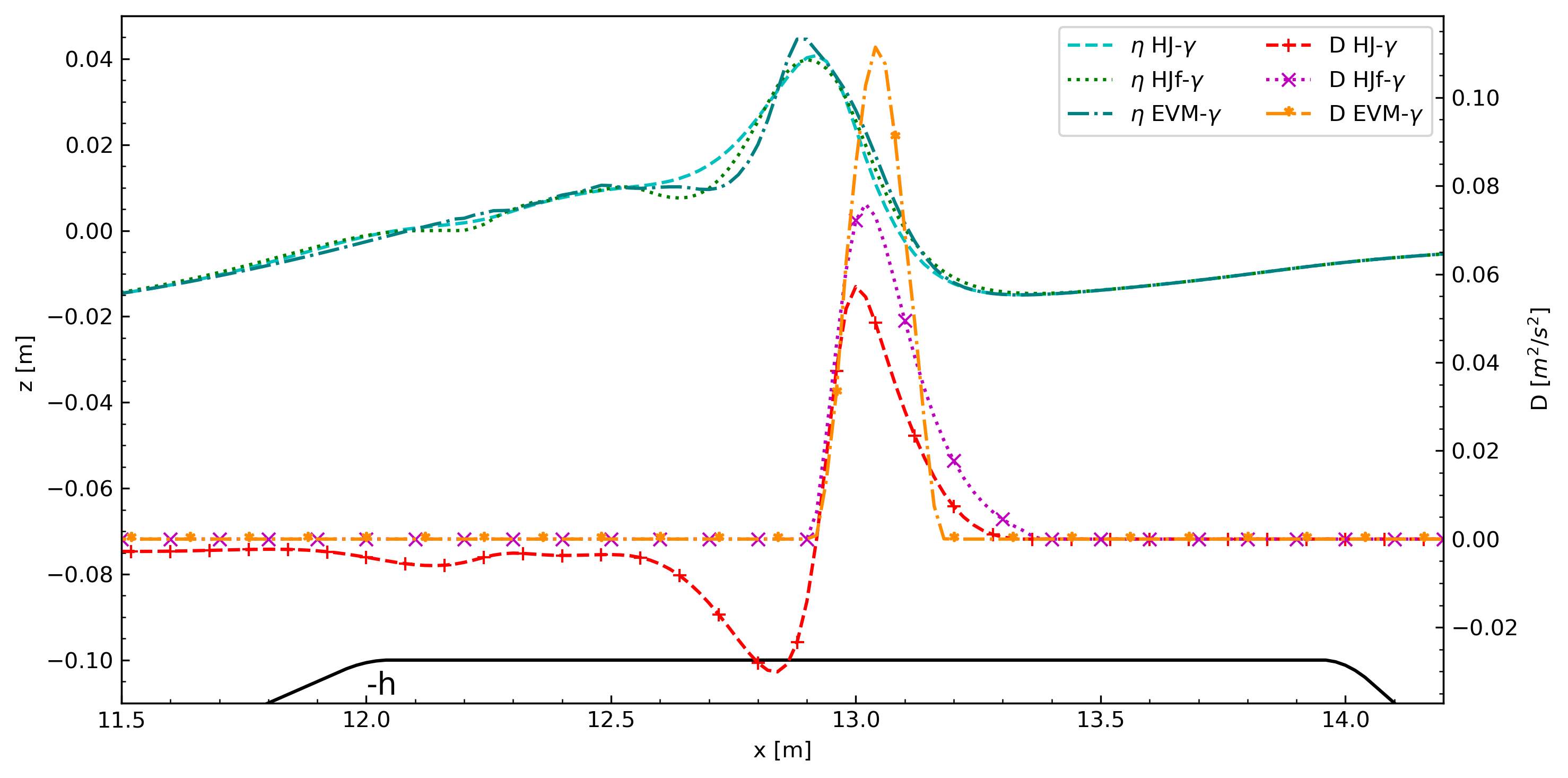}
			\caption{Comparison of the dissipation applied to Eq. \eqref{eq:zak2} for the case of regular wave breaking over a bar presented in section \ref{subsec:BBreg} ($H = 0.054$ m, $T = 2.5$ s). The methods of dissipation HJ, HJf and EVM are presented with the initiation criterion using $\gamma_i$}
			\label{fig:compaD}
		\end{figure}

		The second method (denoted HJf) is a variation of the HJ method, tested by \citet{papoutsellis_modeling_2019} and here.
		It consists in reducing the defined spatial extent of the wave breaking dissipation zone to the wave front, from the wave crest $x_c$ to the wave front $x_f$.
		In order to keep this method stable, a linear transition increasing from $0$ to $1$ over $8$ time steps was used to add progressively in time the term $D=-P_{surf}/\rho$ to Eq. \ref{eq:zak2}.
		This temporal transition reduced the occurrence of instabilities appearing with the abrupt addition of $P_{surf}$,  without impacting significantly the simulated wave energy dissipation.
		\figurename{}~\ref{fig:compaD} shows the spatial extent and variation of $P_{surf}$ for the first identified breaking wave of the simulations in section \ref{subsec:BBreg}.
		\\

		The third dissipation method uses an eddy viscosity model to dissipate energy due to wave breaking.
		This method, denoted EVM, was developed by adding a dissipation term based on the tangential velocity at the surface to the momentum equations in Boussinesq-type models.
		For example, in their model,  \citet{kennedy_boussinesq_2000} compute the dissipation term $R$  as:
		\begin{equation}
			R=\dfrac{1}{h+\eta}\partial_x F, \quad \text{for } x\in [x_c,x_t], \label{eq:Rken}
		\end{equation}
		where $F= - \varDelta \: (h+\eta) \: (\partial_t\eta)^2$.
		Eq. \eqref{eq:Rken} is derived by imposing the additional condition that the total wave momentum should be conserved for waves propagating over a flat bottom.
		The term $\varDelta=\delta^2 B( \partial_t \eta)$ controls the breaking process, and $\delta$ determines the magnitude of the dissipation \citep{kennedy_boussinesq_2000, kurnia_high_2014}.
		$\delta$ is typically chosen in the interval $[0.9, 1.5]$ and calibrated to match experimental measurements.
		The function $B$ depends on $\partial_t\eta$ and varies between $0$ and $1$ in time and space controlling the initiation and termination of the breaking process by replacing $N$ by $\partial_t\eta$ in the formula:
		\begin{equation}
		B(N)=\begin{cases}
			0        & \text{for}\ N\leq N^*      \\
			N/N^*-1, & \text{for}\ N^*<N\leq 2N^* \\
			1        & \text{for}\ N \geq 2N^*.
		\end{cases}
		\end{equation}
		The parameter $N^*$ varies in time from an initial value $N_i$ to a final value $N_f$, following the linear relation:
		\begin{equation}
		N^*=\begin{cases}
			N_i - \dfrac{t-t_i}{T^*} (N_i - N_t) & \text{for}\ t_i\leq t < t_i+T^* \\
			N_t                 & \text{for}\ t\geq t_i+T^*.
		\end{cases}
		\end{equation}
		Here, $t_i$ is the initial time of breaking of the considered wave, and $T^*$ is the transition time.
		The values $N_i$ and $N_t$ depend on the breaking criterion, and \citet{kennedy_boussinesq_2000} proposed using $N_i=\gamma_i\sqrt{gh}$,  $N_t=\gamma_t\sqrt{gh}$ and a transition time $T^*=5\sqrt{g/h}$.
		
		The computed dissipation is only non-zero along the breaking wave between the wave crest and the preceding trough.
		When extending the EVM method to a momentum equation based on potential flow, the added dissipative pressure must satisfy $\partial_x D=R$. 	
		In order compute $D$, Eq. \eqref{eq:Rken} is integrated in space using a two-step Adams-Bashforth method.
		An example of the dissipation computed by EVM is presented	in \figurename{}~\ref{fig:compaD}.
		The spatial extent of the applied dissipation using the EVM method is smaller than those defined using HJ and HJf because of the influence of the function $B$.
		
		When computing the breaking criterion, $u_1/C>\sigma_{i}$, or the dissipation term with the HJ or HJf methods, an accurate estimation of  the wave phase speed $C$ is needed. 
		Here, a partial Hilbert transform  \citep{stansell_experimental_2002, kurnia_high_2014} is used to determine the instantaneous wave number $k$ in the whole domain and to evaluate the crest speed using the linear dispersion relation:
		\begin{equation}
			C(x,t)=\sqrt{ g \dfrac{\tanh (k(x,t)h(x))}{k(x,t)}}. \label{eq:celerity}
		\end{equation}
		To compute $k$, the partial Hilbert transform of the free surface elevation is used with respect to $x$, which is defined by:
		\begin{equation}
			\mathbb{H}[\eta(x,t)]=\dfrac{1}{\pi}p.v.\int_{-\infty}^{+\infty} \dfrac{\eta(x',t)}{x-x'}dx',
		\end{equation}
		where $p.v.$ is the Cauchy principal value of the integral. 
		The Hilbert transform is used to express the phase function, which allows computing the wave number as:
		\begin{equation}
			k(x,t)=\dfrac{1}{\eta^2+\mathbb{H}[\eta]^2} \left( \eta \partial_x\mathbb{H}[\eta] - \mathbb{H}[\eta] \partial_x \eta \right),
		\end{equation}
		in turn used to compute the local wave celerity using Eq. \eqref{eq:celerity}.
					
	\section{Results}
		
	\subsection{General presentation of the simulated experiments}
	
	The implementation of breaking models in whispers3D was tested with four sets of laboratory experiments: regular (\ref{subsec:BBreg}) and then irregular waves (\ref{subsec:BBirreg}), breaking over a submerged bar \citep{beji_experimental_1993}, irregular waves breaking on a plane beach slope \citep{mase_hybrid_1992} (\ref{subsec:MK}), and finally  irregular waves breaking over a submerged sloping bed \citep{adytia_numerical_2018} (\ref{subsec:AHA}).
		
	Details of the simulation parameters, including the time step, $\Delta t$, grid size, $\Delta x$, size of the domain, length of the relaxation and absorption zones, total simulation time, and CPU time are summarized in \tablename{}~\ref{tab:NumSet}. 
	The simulations were run on a laptop equipped with an Intel Core i7-4710MQ ($2.5$ to $3.5$ GHz), $16$ Go DDR3 RAM and a $6$ GB/s SATA3 SSD Drive running on Windows 10.
	The simulations are named after the combination of breaking criterion and dissipation method used, for example: simulation HJ-$\beta$ uses the criterion '$\beta$' (the slope of $\eta$) to activate the dissipation method HJ.
		
	\begin{table}[h!]
		\caption{Parameters used to run the numerical simulations with whispers3D of the four different experimental cases. (N.: Number, L. relax. zone: Length of relaxation zone, L. absor. zone: Length of absorption zone and $T_{simu}$: simulated time)} \label{tab:NumSet}
		\centering
\begin{tabular}{c|cccc}
    \hline
	Bed & bar  & bar  & beach  & slope \\
	Wave type & regular & irregular & irregular & irregular \\
	Case name & SLP & JLP & case 2 & 2009051509 \\
	\hline
	$\Delta x$ (m)  & $0.02$ & $0.02$ & $0.01$ & $0.04$ \\
	$\Delta t$ (s)  & $0.02$ & $0.02$ & $0.01$ & $0.02$ \\
	$x_{\min}$ (m) & $0$  & $-3$  & $-3$   & $9$ \\
	$x_{\max}$ (m) & $25$ & $27$ & $12.9$ &  $45$ \\
	N. of points & $1251$ & $1501$ &  $1591$ &  $901$ \\
	L. relax. zone (m) & $6$ & $9$ & $3$ & $3.225$ \\
	L. absor. zone (m) & $5$ & $9$ & $4$ & $4$ \\
	$T_{simu}$ (s) & 200 & 1200 & 460  & 490 \\
	CPU time (s)   & 600 & 3000 & 3000 & 800 \\
	\hline
\end{tabular}
	\end{table}
		
	Results are presented by comparing time series of the free surface elevation as shown in \figurename{}~\ref{fig:regTSBB}, \ref{fig:irregTSBB}, \ref{fig:TSMK} and \ref{fig:TSAHA}.
	Wave characteristics are presented and compared in \figurename{}~\ref{fig:regPropBB}, \ref{fig:irregPropBB}, \ref{fig:PropMK} and \ref{fig:PropAHA}, including: 
	$H_s$ the significant wave height, $Sk$ the skewness, $Kurt$ the kurtosis, and $As$ the asymmetry, which are computed as:
	\begin{align}
		H_s &= 4  \left\langle \left( \eta - \left\langle \eta \right\rangle  \right)^2 \right\rangle ^{1/2}, \\
		Sk  &= \left. \left\langle \left( \eta - \left\langle \eta \right\rangle  \right)^3 \right\rangle \middle/\left\langle \left( \eta - \left\langle \eta \right\rangle  \right)^2 \right\rangle^{3/2} \right. , \\
		Kurt&=\left. \left\langle \left( \eta - \left\langle \eta \right\rangle   \right)^4 \right\rangle \middle/\left\langle \left( \eta - \left\langle \eta \right\rangle   \right)^2 \right\rangle^{2} \right. -3, \\
		As  &=\left. \left\langle \mathbb{H}\left( \eta - \left\langle \eta \right\rangle   \right)^3 \right\rangle \middle/\left\langle \left( \eta - \left\langle \eta \right\rangle   \right)^2 \right\rangle^{3/2} \right..
	\end{align}
	where $\langle . \rangle$ is the time averaging operator.
	
	The simulations are compared with the experiments by computing the correlation and the normalized root-mean-square deviation, NRMSD, of the free surface elevation time series at the location of the wave gauges.
	The correlation is calculated as the covariance of the two time series divided by the product of their standard deviations (Pearson correlation).
	The correlation (corr) and the NRMSD between signals from simulation (X) and experiment (Y) are computed as:
	\begin{align}
		\text{corr}(X,Y)&=\dfrac{\sum_{t=1}^{T} \left(X_t- \left\langle X \right\rangle \right) \left(Y_t - \left\langle Y \right\rangle \right)}{ \sqrt{ \sum_{t=1}^{T} \left(X_t- \left\langle X \right\rangle \right)^2 \sum_{t=1}^{T} \left(Y_t- \left\langle Y \right\rangle \right)^2 }  }, \\
		\text{NRMSD}(X,Y)&=  \sqrt{\dfrac{\sum_{t=1}^{T} \left( X_t - Y_t \right)^2 }{T}}\dfrac{1}{\max{Y}-\min{Y}} .
	\end{align}
	A deviation of the correlation from the maximal value of $1$ indicates differences in the phase and shape of the wave signals.
	The NRMSD is normalized by the difference between the maximum and minimum values of the experimental time series of the free surface elevation.
	
	Regarding the other wave properties, an increase in skewness indicates a narrowing of wave crests and a widening and flattening of wave troughs.
	Wave asymmetry reaches higher values when the front of the wave shortens and steepens, and the back of the wave therefore lengthens.
	The kurtosis is a measure of the peakedness of the free surface elevation distribution.
	Its value is $0$ for a Gaussian distribution.
	When the free surface position is represented by a sinusoidal function, the distribution has a kurtosis of $-1.5$.
	Values larger than $0$ indicate that the signal has tall, thin waves.
	
	Finally, for the regular wave case, the amplitude of the first six harmonics are presented in \figurename{}~\ref{fig:regFFTBB} as a function of the distance, whereas, for the irregular wave cases, wave amplitude spectra are presented in \figurename{}~\ref{fig:irregFFTBB}, \ref{fig:FFTMK}, and \ref{fig:FFTAHA} where the frequency on the horizontal axis being normalized by the peak frequency specified in the experiments.
	
	\subsection{Regular waves breaking over a submerged bar} \label{subsec:BBreg}
\begin{figure}[h!]
		\centering
		\includegraphics[width=\linewidth]{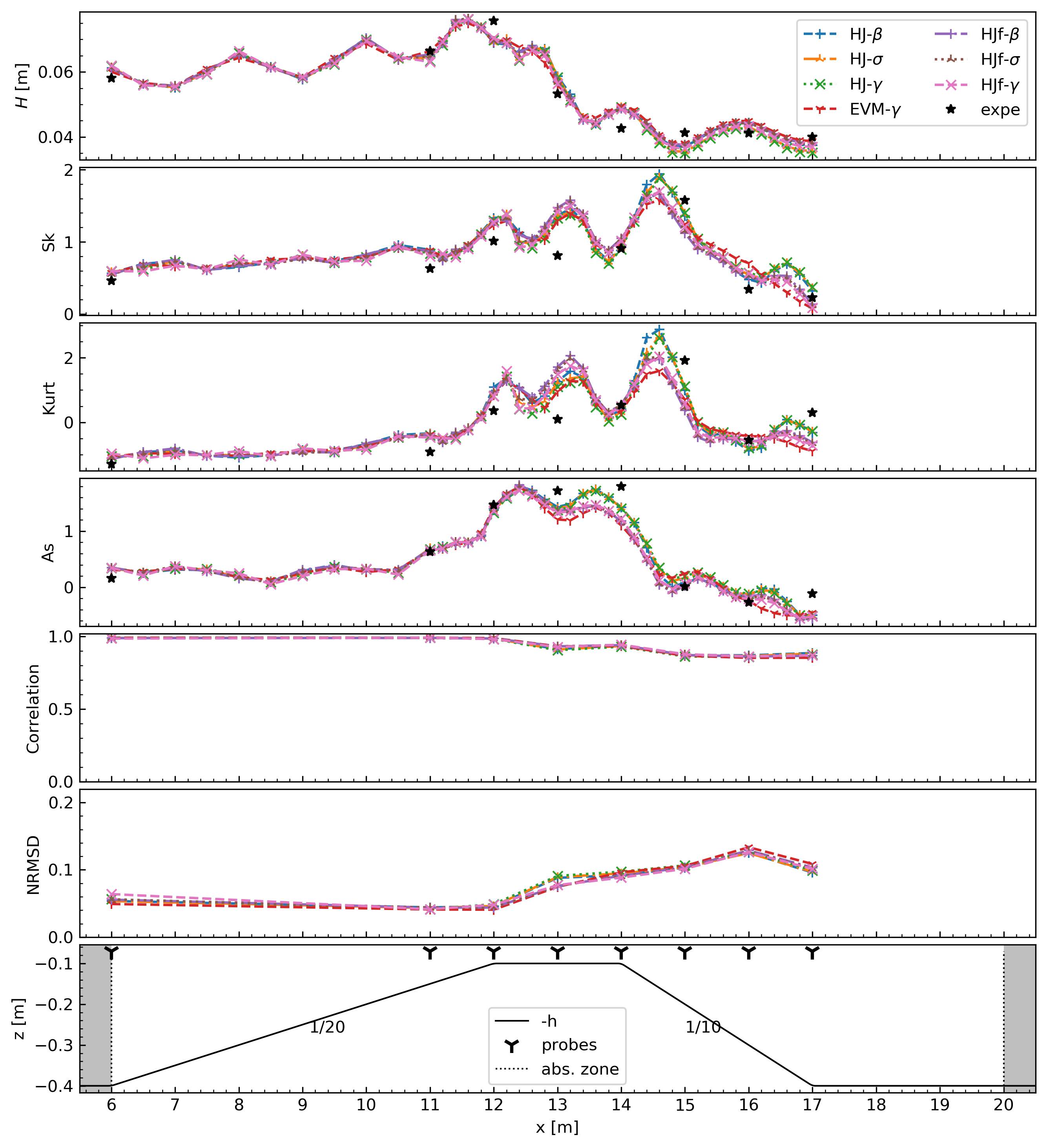}
		\caption{Comparison of the wave properties between the experimental data \citep{beji_experimental_1993} and numerical simulation results for regular waves breaking over a bar ($H = 0.054$ m, $T = 2.5$ s)}
		\label{fig:regPropBB}
	\end{figure}

	\citet{beji_experimental_1993} conducted several experiments of wave propagation over a trapezoidal bar.
	The first test is focused on experiments propagating long waves (wave height $H = 0.054$ m and period $T = 2.5$ s), resulting in plunging wave breaking \citep{beji_experimental_1993}.
	The bar varies in depth from $0.4$ m to $0.1$ m (See bottom profile and location of probes in the lower panel of \figurename{}~\ref{fig:regPropBB}), with a $1$:$20$ upslope and a $1$:$10$ downslope starting at $x=6$ m and $x=14$ m, respectively.
	The free surface was measured with $8$ wave gauges located at $6$, $11$, $12$, $13$, $14$, $15$, $16$ and $17$ m.
	In the numerical simulations, the incident waves were generated using the signal recorded at the first probe.	
	\tablename{} \ref{tab:BBref} summarizes the combination of wave breaking criterion and dissipation mechanisms used in the model, as well as the parameters used to initiate and terminate wave breaking, and the dissipation coefficient.
		
	\begin{table}[h!]
		\caption{Wave breaking parameters for the case of regular waves propagating over a submerged bar \citep{beji_experimental_1993}}\label{tab:BBref}
		\centering
		\begin{tabular}{ r|ccccccc}
    \hline
	Case name  & HJ-$\beta$ & HJ-$\gamma$ & HJ-$\sigma$ & HJf-$\beta$ & HJf-$\gamma$ & HJf-$\sigma$ & EVM-$\gamma$ \\
	Strength coef. & $\mu=1$     & $\mu=1$         & $\mu=1$        & $\mu=1$        & $\mu=1$         &  $\mu=1$       & $\delta=1.2$    \\
	Initiation  & $\beta_{i}=35^\circ$ & $\gamma_i=0.6$  & $\sigma_i=0.84$    & $\beta_{i}=35^\circ$ & $\gamma_i=0.6$  & $\sigma_i=0.84$    &   $\gamma_i=0.6$  \\
	Cessation   & $\beta_{t}=12^\circ$ & $\gamma_t=0.15$ & $\beta_{t}=12^\circ$ & $\beta_{t}=10^\circ$ & $\gamma_t=0.15$ & $\beta_{t}=10^\circ$ &     $\gamma_t=0.11$ \\
	\hline
\end{tabular}
	\end{table}

	\begin{figure}[h!]
		\centering
		\includegraphics[width=\linewidth]{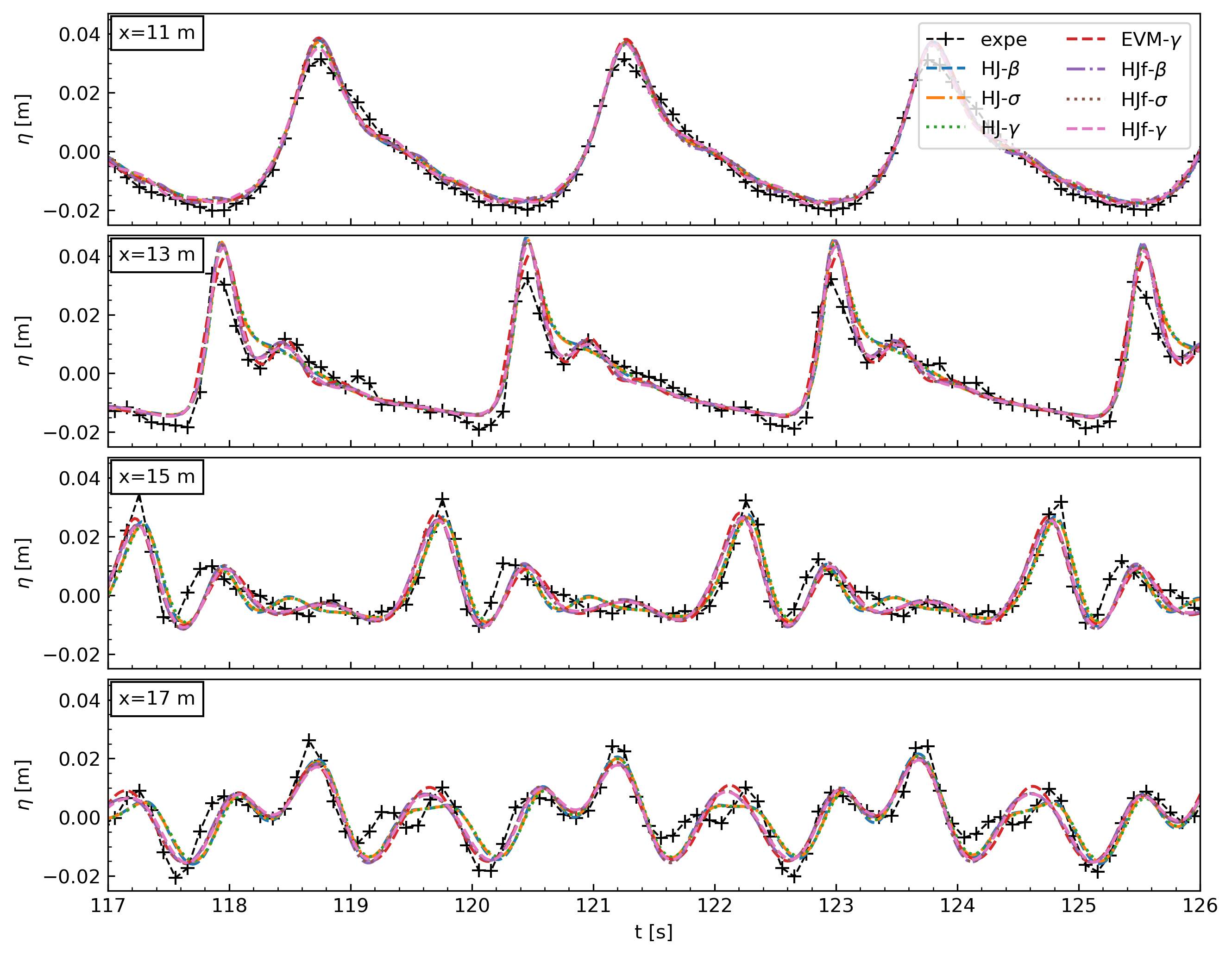}
		\caption{Time series of the free surface elevation for regular waves ($H = 0.054$ m, $T = 2.5$ s) breaking over a trapezoidal bar. Comparison between the experiments \citep{beji_experimental_1993} and numerical simulations using different combinations of breaking criteria and dissipation methods}
		\label{fig:regTSBB}
	\end{figure}
	
	\begin{figure}[h!]
		\centering
		\includegraphics[width=\linewidth]{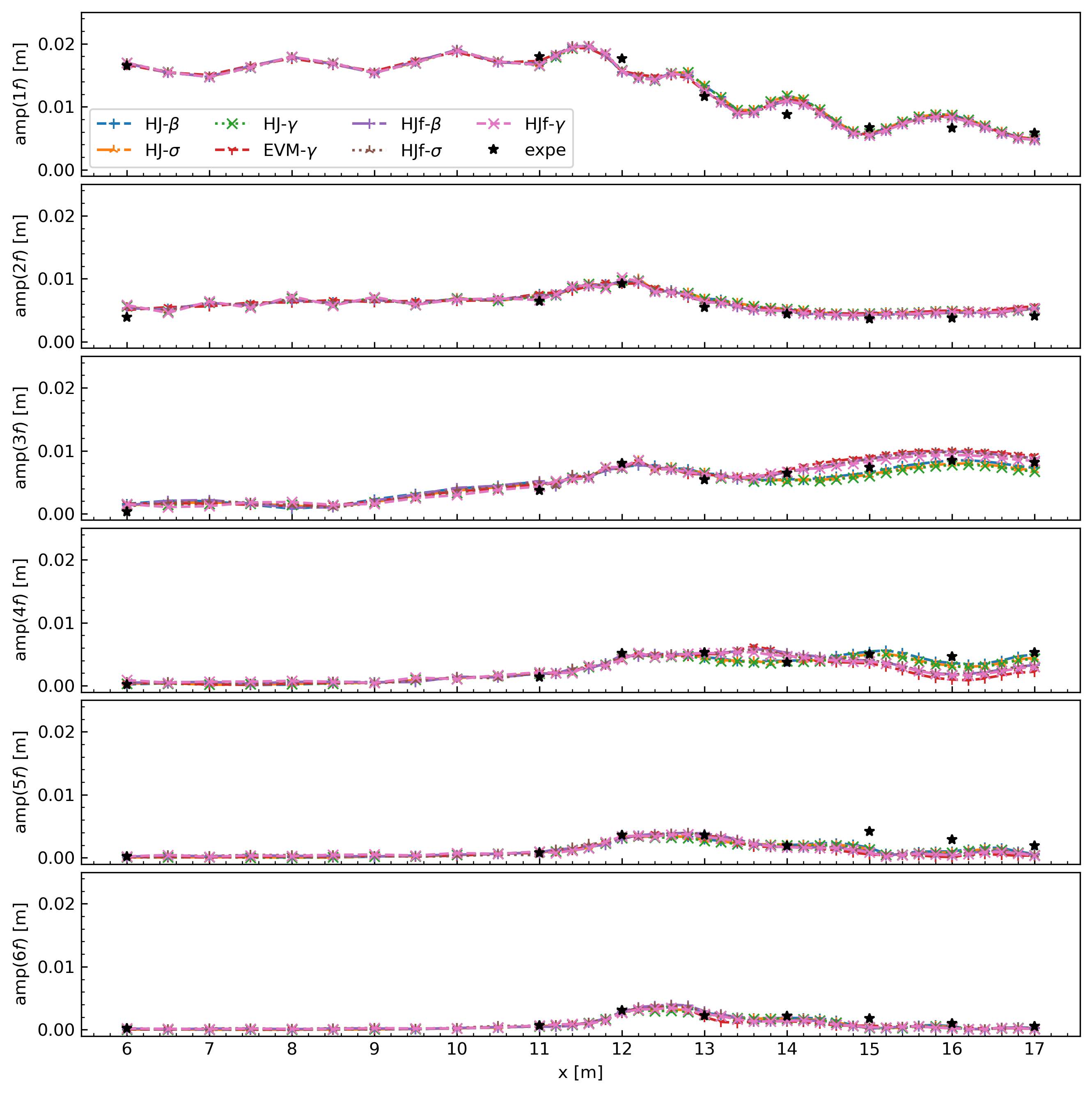}
		\caption{Comparison of the spatial evolution of the first six harmonic amplitudes (at frequencies $1f$, $2f$,…, $6f$) of the free surface elevation between the experimental data \citep{beji_experimental_1993} and numerical simulation results for regular waves breaking over a bar ($H = 0.054$ m, $T = 2.5$ s)}
		\label{fig:regFFTBB}
	\end{figure}

	In the simulations and experiments, as the waves travel upslope, the wave height increases during the shoaling process.
	The wave fronts shorten and steepen, with the wave tail becoming longer (\figurename{}~\ref{fig:regTSBB}), which results in an increase in asymmetry before $x=13$ m (\figurename{}~\ref{fig:regPropBB}).
	Wave breaking occurs over the top of the bar. 
	In the simulations, breaking is initiated between $x=12.17$ and $12.20$ m and terminates between $x=14.15$ and $14.57$ m, depending on the combination of parameters used (\tablename{} \ref{tab:BBref}).
	As waves break and the wave energy is dissipated, the wave height decreases, and the waves decompose into a series of free waves of higher frequencies in the deeper part of the domain (\figurename{}~\ref{fig:regTSBB} and \ref{fig:regFFTBB}).
	
	For the selected parameters, the wave evolution is nearly identical between all simulations (\figurename{} \ref{fig:regTSBB}). 
	The time traces and wave properties resulting from simulations with the same dissipation method but different breaking criteria are almost superimposed. 
	The activation of the dissipation occurs at the same time for each wave since this test case simulates regular waves.
	Moreover, using EVM or HJf produces similar results since the applied dissipation mechanisms have similar values and are both applied on the wave front (\figurename{} \ref{fig:compaD}), whereas the simulations using HJ produce time series with slightly smaller secondary oscillations ($x=13$ m, \figurename{} \ref{fig:regTSBB}).

	Overall, the simulation results agree well with the experiments.
	In the simulations, the wave crest height is slightly overestimated before breaking and underestimated after breaking (\figurename{} \ref{fig:regTSBB}). 
	Nonetheless, the wave height evolution agrees well with the measurements (\figurename{} \ref{fig:regPropBB}).
	\figurename{} \ref{fig:regPropBB} shows a high correlation between the simulated and observed free surface position with a minimal  correlation of $85\%$ and a NRMSD less than $0.14$.
	When comparing to results obtained in previous studies, as an example, the correlation with the experiments at $x=15$ m in \citet{kurnia_high_2014} is $0.863$, while here, the correlation for simulation HJf-$\gamma$ is $0.871$. 
	Finally, the decomposition into higher harmonics is well captured in all approaches as shown in \figurename{} \ref{fig:regFFTBB}, showing the spatial evolution of the amplitudes of the first six harmonics of the wave signal.
	The amplitude of the fundamental harmonic decreases as the waves break. 
	Good agreement between the simulated and measured harmonic amplitudes is obtained, even for the higher harmonics, although small differences can be observed after $x=13$ m for harmonics $3$ to $5$, depending on the proposed breaking method.
	
	The  results here are in agreement with those obtained by \citet{papoutsellis_modeling_2019} using similar wave breaking parameterization approaches but with a different technique to solve the DtN problem.
	This test case confirms the ability of the proposed methods to simulate regular breaking waves.
	
	\clearpage
	
	\subsection{Irregular waves breaking over a submerged bar} \label{subsec:BBirreg}
\begin{figure}[h!]
		\centering
		\includegraphics[ width=\linewidth]{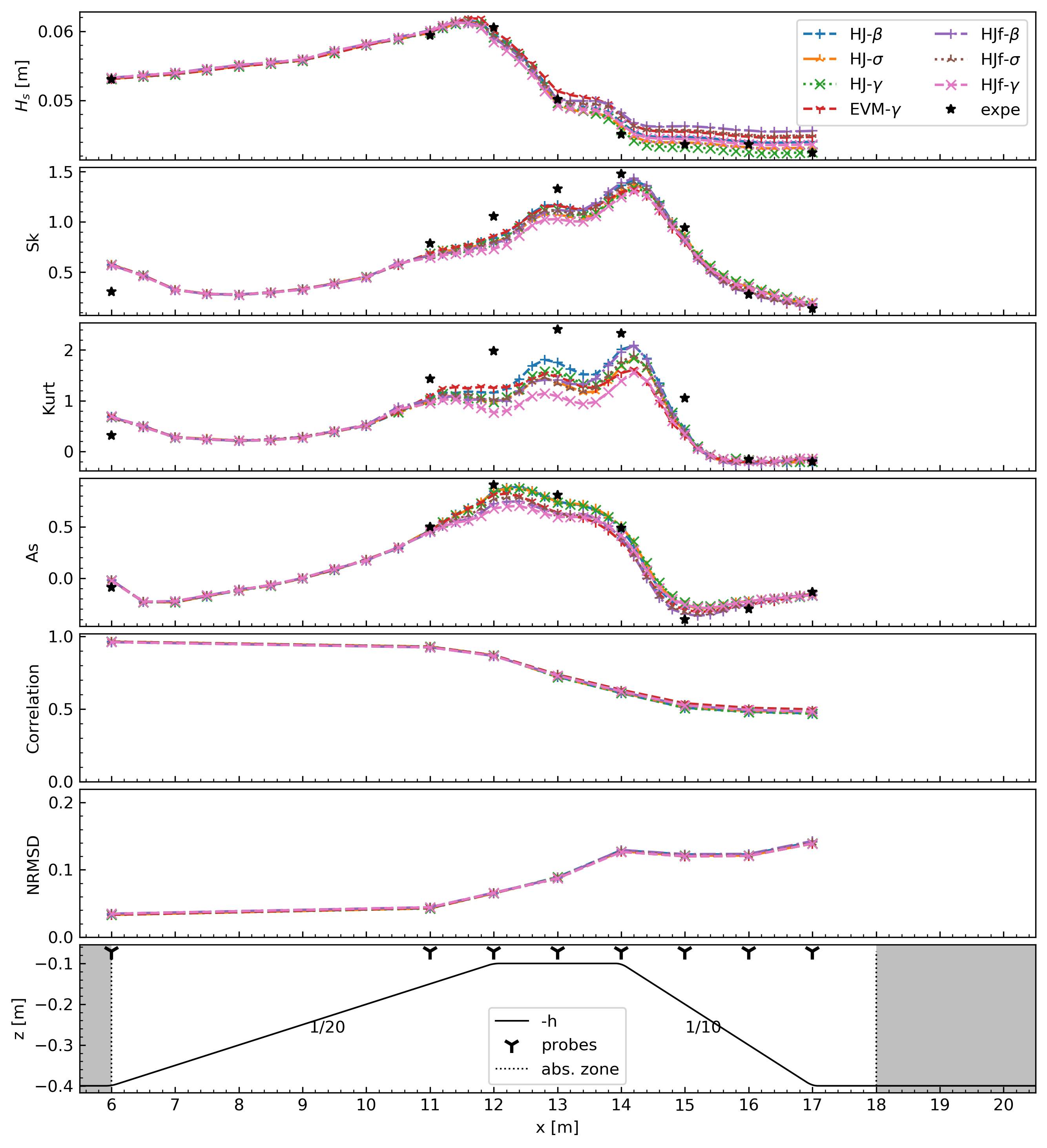}
		\caption{Comparison of the wave properties of the experimental data \citep{beji_experimental_1993} and numerical simulation results for a test with irregular waves breaking over a bar ($H_s = 0.049$~m, $T_p = 2.5$ s)}
		\label{fig:irregPropBB}
	\end{figure}

	Using the same bathymetry presented in the previous section (see lower panel of \figurename{} \ref{fig:irregPropBB}), simulations were performed for irregular waves characterized by a JONSWAP-type spectrum having a significant wave height $H_s = 0.049$~m and a peak period $T_p = 2.5$~s \citep{beji_experimental_1993}.
	\tablename{} \ref{tab:irregBBref} summarizes the combination of wave breaking criterion and dissipation mechanism used in the model, as well as the parameters used to initiate and terminate wave breaking, and the dissipation coefficient.
	
	\begin{table}[h!]
		\caption{Wave breaking parameters for the case of irregular waves propagating over a submerged bar \citep{beji_experimental_1993}}\label{tab:irregBBref}
		\centering
		\begin{tabular}{ r|ccccccc}
    \hline
	Case name  & HJ-$\beta$ & HJ-$\gamma$ & HJ-$\sigma$ & HJf-$\beta$ & HJf-$\gamma$ & HJf-$\sigma$ & EVM-$\gamma$t \\
	Strength coef. & $\mu=1$     & $\mu=1$         & $\mu=1$        & $\mu=1.1$        & $\mu=1.2$         &  $\mu=1$       & $\delta=1.2$    \\
	Initiation  & $\beta_{i}=35^\circ$ & $\gamma_i=0.65$  & $\sigma_i=0.84$    & $\beta_{i}=30^\circ$ & $\gamma_i=0.55$  & $\sigma_i=0.84$    &   $\gamma_i=0.6$  \\
	Cessation   & $\beta_{t}=12^\circ$ & $\gamma_t=0.18$ & $\beta_{t}=12^\circ$ & $\beta_{t}=12^\circ$ & $\gamma_t=0.18$ & $\beta_{t}=12^\circ$ &     $\gamma_t=0.15$ \\
	\hline
\end{tabular}
	\end{table}
	
\begin{figure}[h!]
		\centering
		\includegraphics[width=\linewidth]{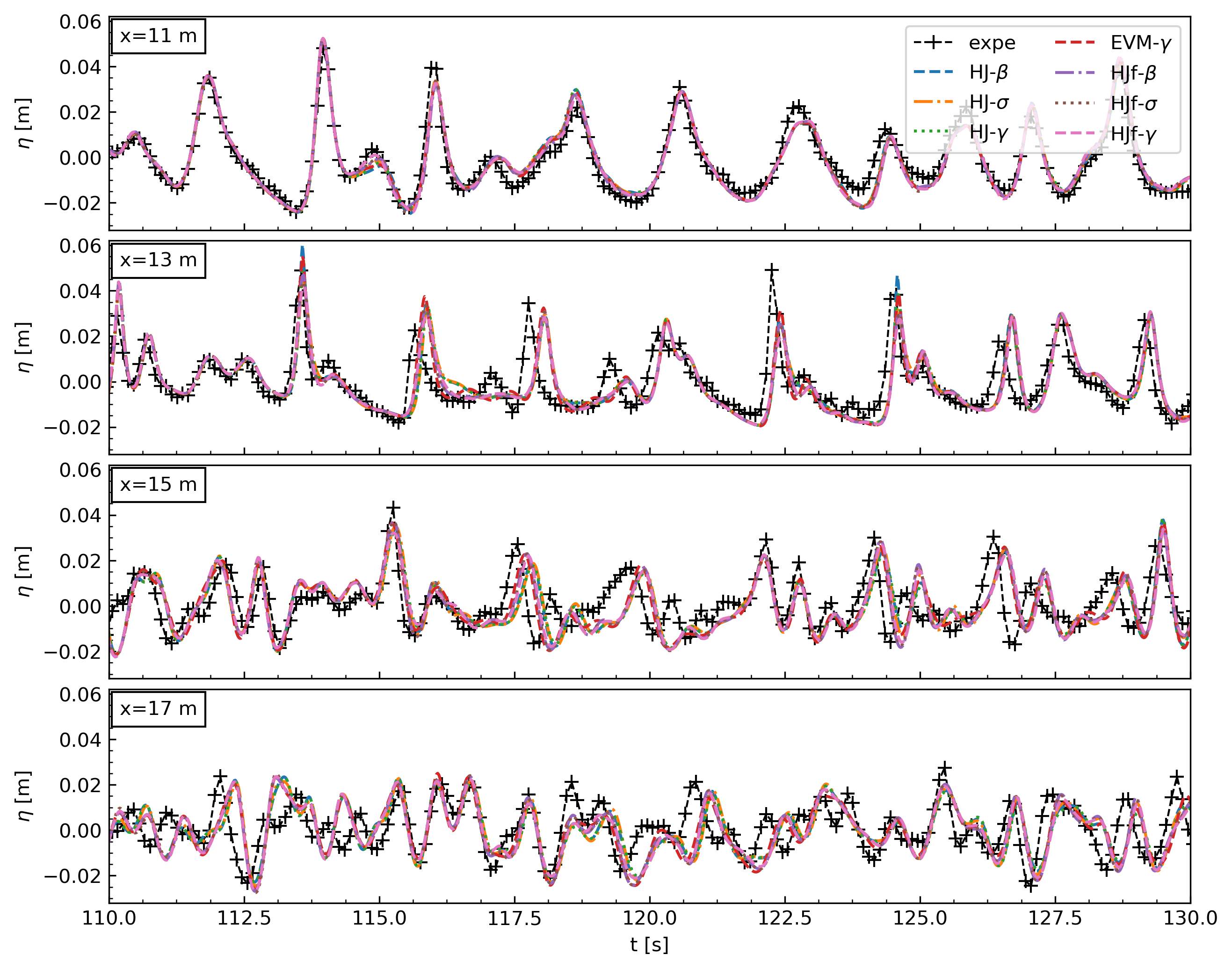}
		\caption{Time series of the free surface elevation for irregular waves ($H_s = 0.049$~m, $T_p = 2.5$ s) breaking over a trapezoidal bar. Comparison between the experiments \citep{beji_experimental_1993} and numerical simulations using different combinations of breaking criteria and dissipation methods}
		\label{fig:irregTSBB}
	\end{figure}

	In the simulations and experiments, as the waves travel upslope, the wave crests narrow, and the waves lose their symmetry (\figurename{} \ref{fig:irregTSBB}). 
	During the simulations, wave breaking is initiated between $10.5$ and $11$ m for a few waves, with on average, dissipation starting at $x\approx12.4$ m.
	The dissipation terminates, on average, after $x \approx 14.4$ m.
	The wave asymmetry increases until $x \approx 12$ m, and decreases when the waves start breaking, whereas the skewness decreases only over the downslope (\figurename{} \ref{fig:irregPropBB}).
	
\begin{figure}[h!p]
		\centering
		\includegraphics[width=\linewidth]{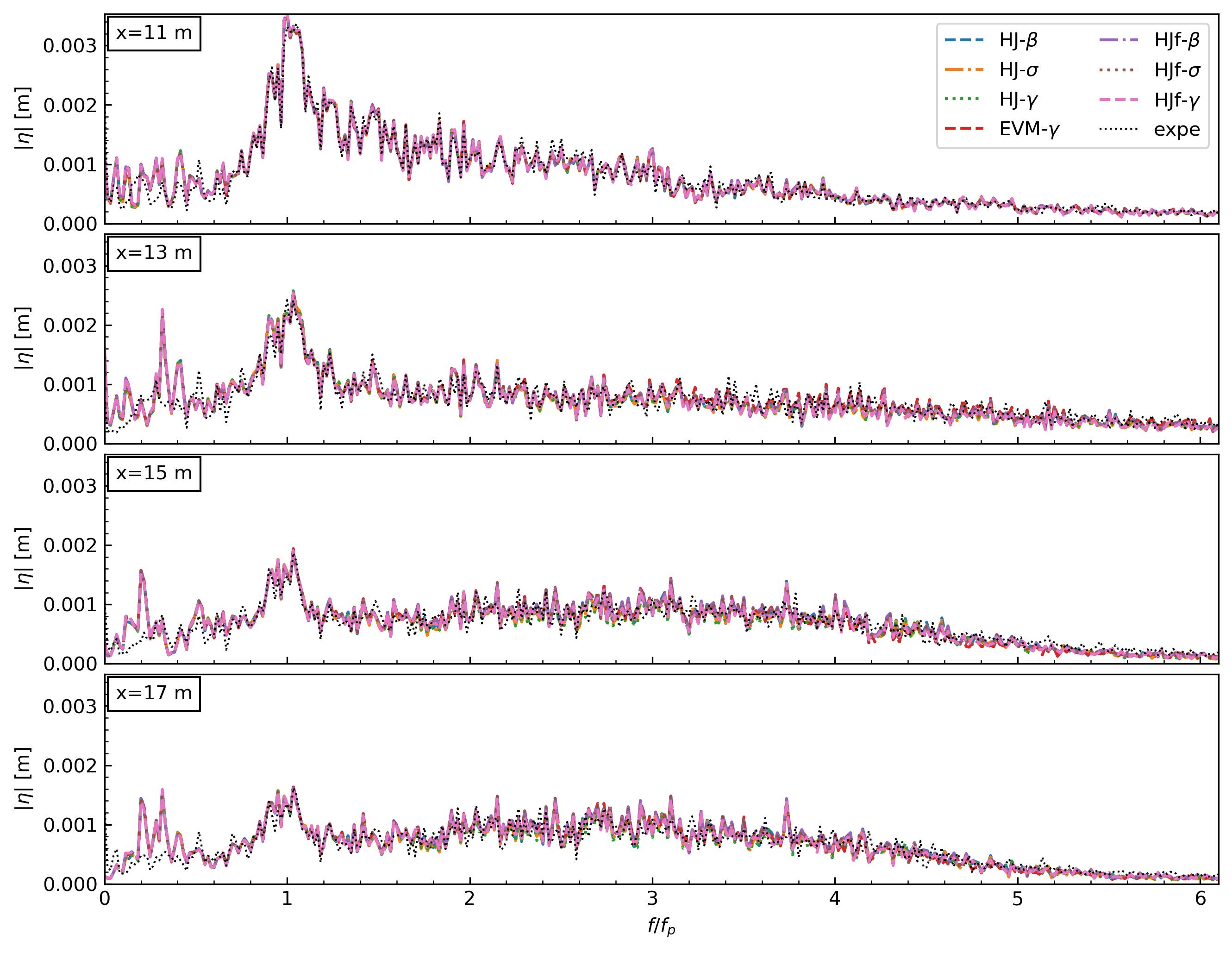}
		\caption{Comparison of the wave amplitude spectra between the experimental data \citep{beji_experimental_1993} and numerical simulation results for irregular waves breaking over a bar ($H_s = 0.049$~m, $T_p = 2.5$ s)}
		\label{fig:irregFFTBB}
	\end{figure}

	Similar to the regular wave case, the wave height decreases after wave breaking is initiated, and the incident waves decompose into higher frequency components while they travel over the downslope  (\figurename{} \ref{fig:irregFFTBB}). 

	Overall, the simulation results are similar. 
	Differences appear over the top of the bar for the asymmetry and kurtosis, as well as the wave height over the downslope.
	At the end of the downslope ($x=17$ m), a $3$ mm difference between the wave heights of simulations HJ-$\gamma$ and HJf-$\beta$ is observed (\figurename{} \ref{fig:irregPropBB}).
	The selection of values for the breaking initiation and termination and power of the dissipation influences the differences observed between the simulations (\tablename{}~\ref{tab:irregBBref}).
	Moreover, the refinement of the breaking parameters for one statistical wave parameter can be detrimental to another statistical parameter.
	For instance, the results from HJ-$\beta$ slightly overestimate the wave height compared to the experiments but provide a better match of the observed skewness and the asymmetry (\figurename{} \ref{fig:irregPropBB}).
	
	Overall, the simulations results agree with the experimental measurements (\figurename{} \ref{fig:irregPropBB}).
	The skewness and kurtosis are slightly underestimated over the top of the bar, but the numerical results match the experimental measurements over the downslope.
	Cases HJ-$\sigma$, HJ-$\gamma$ and HJf-$\gamma$ show the best agreement with the experiments with respect to the wave height, whereas HJ-$\beta$ shows the best agreement for the other wave properties (\figurename{} \ref{fig:irregPropBB}). 
	This is a consequence of the difficulties in optimizing the $3$ parameters controlling wave breaking.
	In \figurename{} \ref{fig:irregTSBB} a few waves appear to be out of phase.
	As a consequence, the correlation of the simulated and measured time series decreases to reach a minimal value of $47\%$ for HJ-$\gamma$.
	However, the NRMSD remains less than $0.14$ and, the amplitude spectra are in good agreement with the experiments over most of the frequency range (\figurename{}~\ref{fig:irregFFTBB}), except for the low frequency range ($f < 0.5 f_p$) where the presence of some peaks in the simulation results can be attributed to the incomplete absorption of long waves in  the relaxation zones.
	
	This case confirms the capacity of the proposed methods to simulate irregular breaking waves in the presence of strong dispersion and nonlinearity.
	
	\clearpage
	
	\subsection{Irregular waves breaking on a plane beach} \label{subsec:MK}
	
\begin{figure}[h!]
		\centering
		\includegraphics[ width=\linewidth]{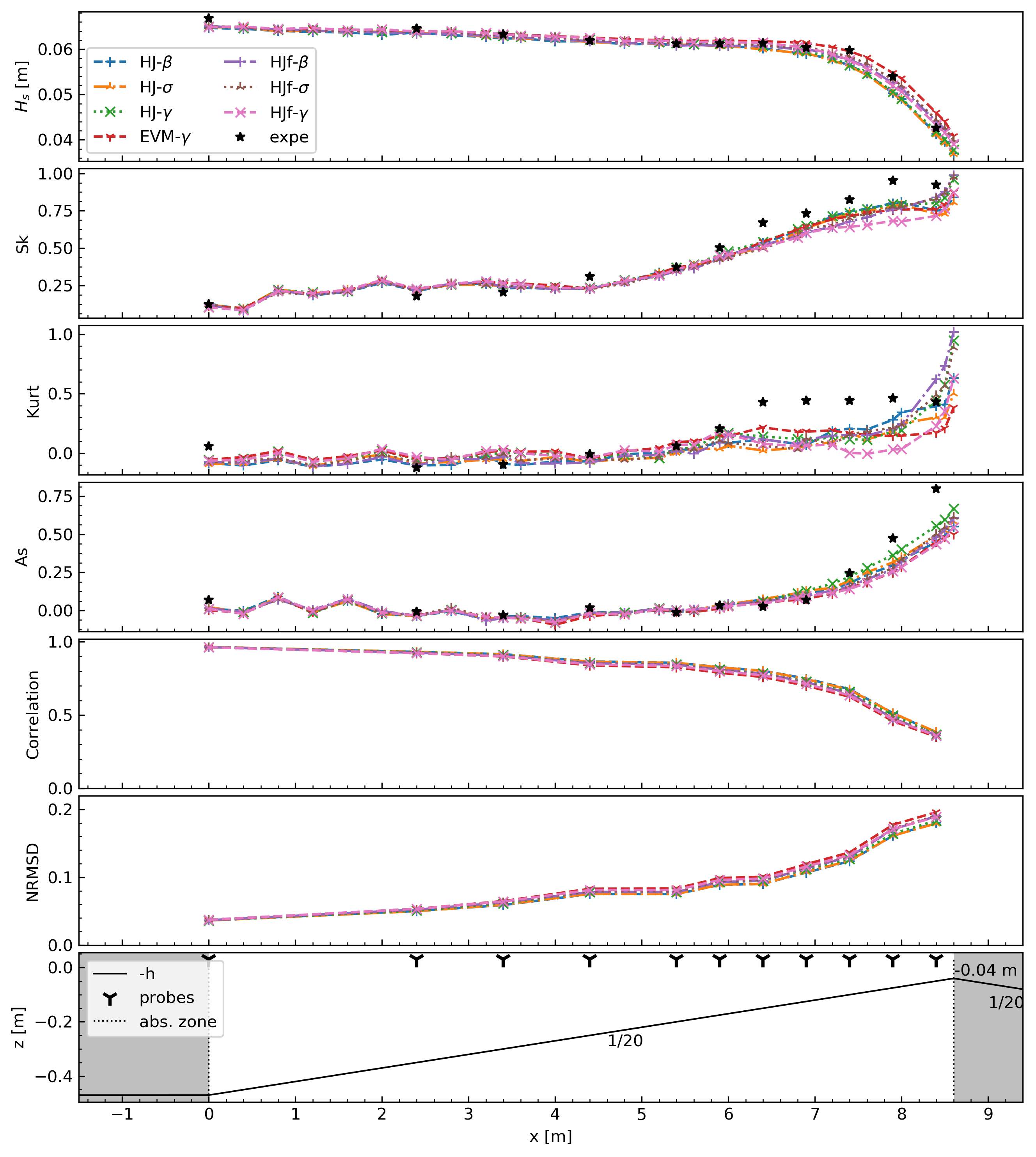}
		\caption{Comparison of the wave properties between the experimental data \citep{mase_hybrid_1992} and numerical simulation results for irregular waves breaking over a beach ($H_s = 0.06$ m, $T_p = 1$ s)}
		\label{fig:PropMK}
	\end{figure}

	The third case is the breaking of irregular waves on a plane beach of slope $1$:$20$ following the  experiments of \citet{mase_hybrid_1992}.
	The waves propagate for $10$ m on a flat bed with a $0.47$ m water depth before propagating up a sloped beach starting at $x=0$ m.
	The free surface was measured by $12$ probes located at $0$, $2.4$, $3.4$, $4.4$, $5.4$, $5.9$, $6.4$, $6.9$, $7.4$, $7.9$, $8.4$, $8.9$ m.
	In the simulations, the beach is modified due to the absence of run-up in the code.
	In the numerical domain, when the depth reaches $0.04$ m, the bathymetry deepens again (see lower panel of \figurename{}~\ref{fig:PropMK}).
	The absorption layer starts at the location of the peak of the bathymetry (\tablename{} \ref{tab:NumSet}).
	The generated waves correspond to a Pierson-Moskowitz spectrum with a peak frequency $f_p=1$ Hz and a significant wave height $H_s=0.06$ m.
	The combination of wave breaking criteria and dissipation mechanisms used for the $7$ simulations are summarized in \tablename{} \ref{tab:MKref}.
	
	\begin{table}[h!]
		\caption{Wave breaking parameters for the case of irregular waves propagating on a plane beach \citep{mase_hybrid_1992}}\label{tab:MKref}
		\centering
		\begin{tabular}{ r|ccccccc}
    \hline
	Case name  & HJ-$\beta$ & HJ-$\gamma$ & HJ-$\sigma$ & HJf-$\beta$ & HJf-$\gamma$ & HJf-$\sigma$ & EVM-$\gamma$ \\
	Strength coef. & $\mu=1.2$     & $\mu=0.9$         & $\mu=1.2$        & $\mu=0.9$        & $\mu=0.9$         &  $\mu=0.9$       & $\delta=1.1$    \\
	Initiation  & $\beta_{i}=30^\circ$ & $\gamma_i=0.5$  & $\sigma_i=0.84$    & $\beta_{i}=30^\circ$ & $\gamma_i=0.5$  & $\sigma_i=0.84$    &   $\gamma_i=0.5$  \\
	Cessation   & $\beta_{t}=15^\circ$ & $\gamma_t=0.15$ & $\beta_{t}=15^\circ$ & $\beta_{t}=10^\circ$ & $\gamma_t=0.15$ & $\beta_{t}=14^\circ$ &     $\gamma_t=0.15$ \\
	\hline
\end{tabular}
	\end{table}
		
\begin{figure}[h!]
		\centering
		\includegraphics[width=\linewidth]{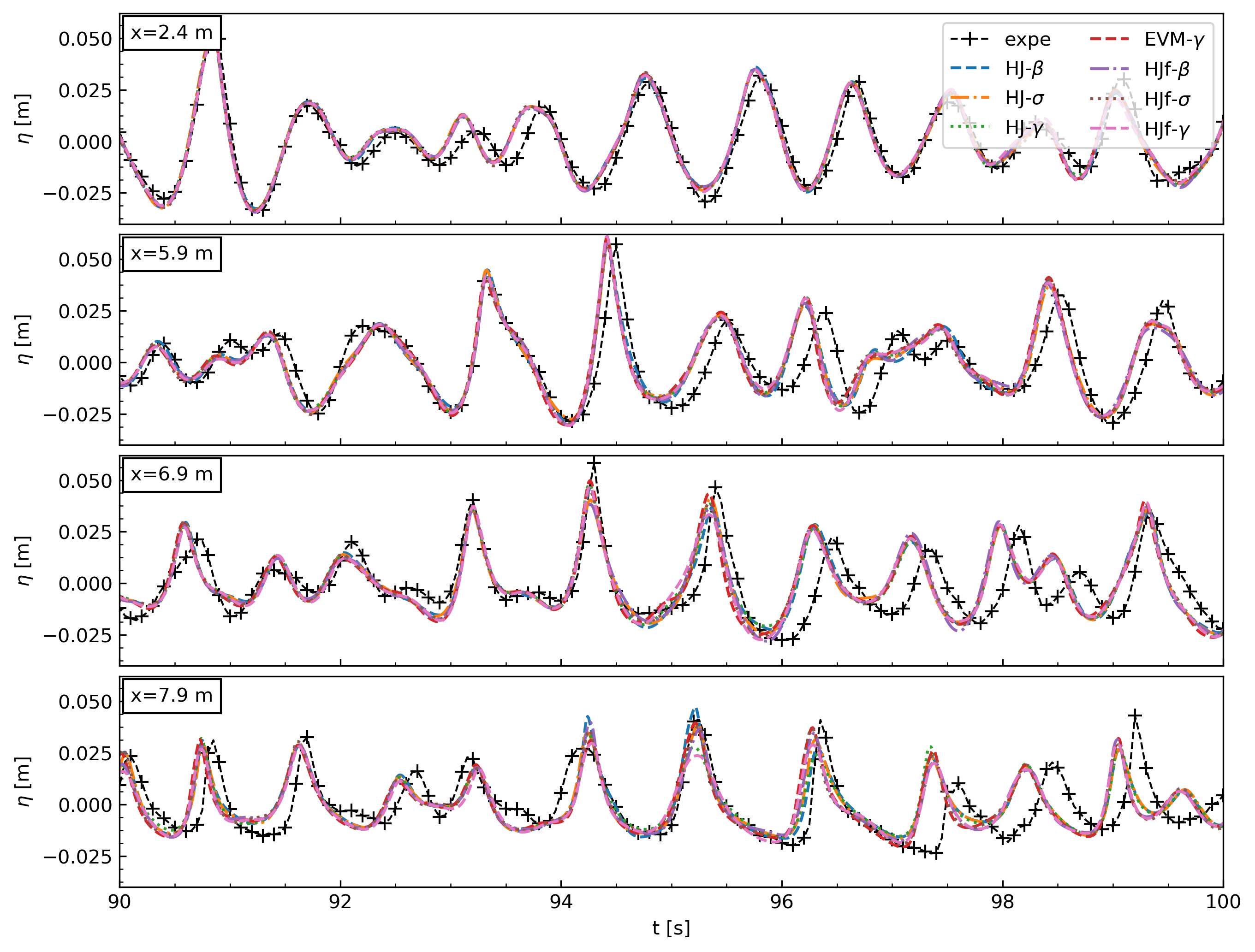}
		\caption{Time series of the free surface elevation for irregular waves breaking over a beach ($H_s = 0.06$ m, $T_p = 1$ s). Comparison between the experiments \citep{mase_hybrid_1992} and numerical simulations using different combinations of breaking criteria and dissipation methods}
		\label{fig:TSMK}
	\end{figure}

\begin{figure}[h!]
		\centering
		\includegraphics[width=\linewidth]{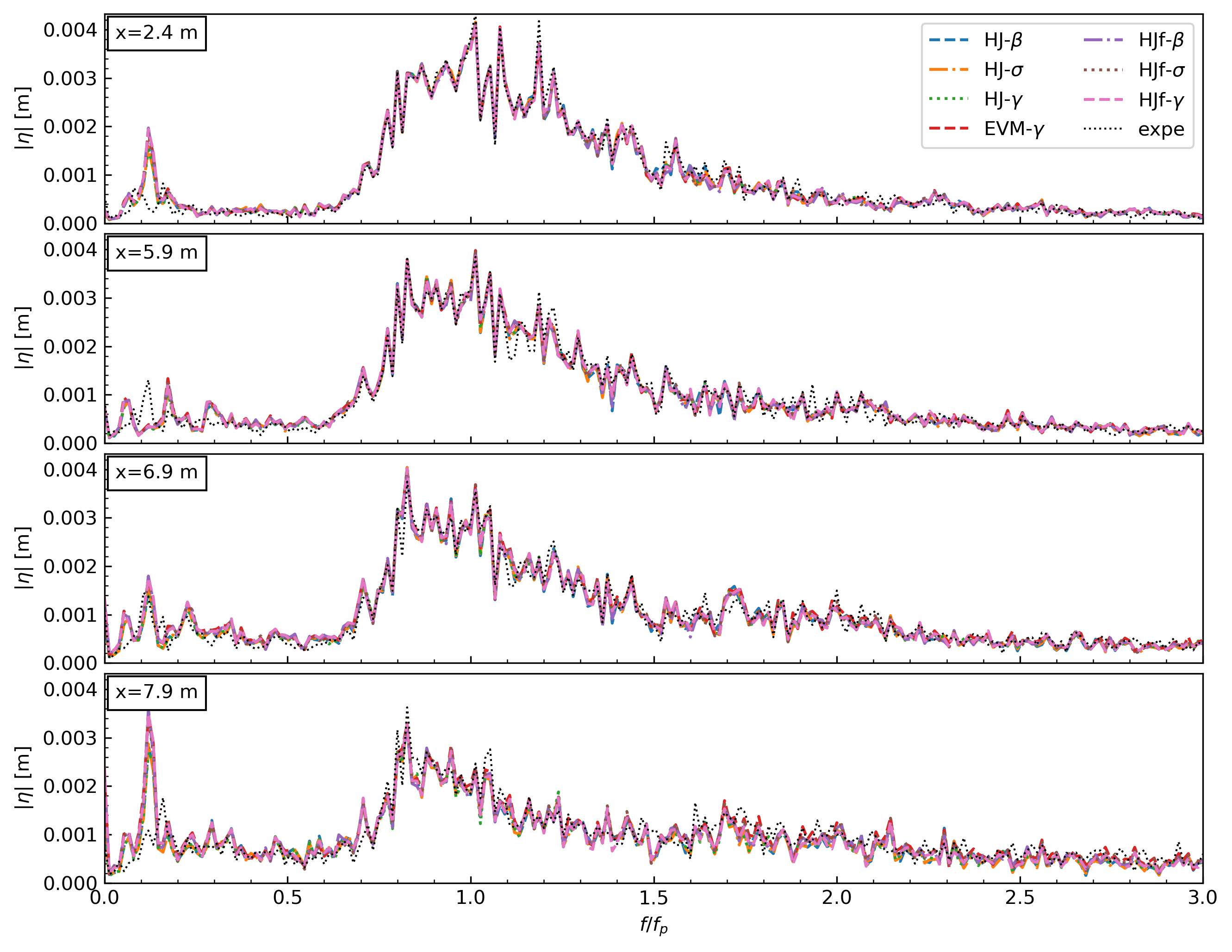}
		\caption{Comparison of the wave amplitude spectra between the experimental data \citep{mase_hybrid_1992} and numerical simulation results for irregular waves breaking over a beach ($H_s = 0.06$ m, $T_p = 1$ s)}
		\label{fig:FFTMK}
	\end{figure}

	In the simulations and experiments, the same changes in wave shape are observed. 
	As the waves travel upslope, they become progressively more asymmetrical, with narrower crests (\figurename{} \ref{fig:TSMK} and \ref{fig:PropMK}).
	In the simulations, for most waves, the dissipation is activated after $x=7$ m.
	As the water depth decreases, \figurename{} \ref{fig:FFTMK} shows a decrease in energy around the peak frequency, as well as the transfer of energy from the peak frequency toward lower and higher frequencies.
	
	Comparing the simulation results obtained by the different methods, the numerical results are again similar, and the simulated time series are nearly superimposed (\figurename{} \ref{fig:TSMK}).
	Differences are only visible when focusing on individual waves where the maximum or minimum values of the wave crests and troughs vary, while the phase remains nearly the same.
	The wave spectra are also similar for each simulation with the curves nearly superimposed (\figurename{} \ref{fig:FFTMK}). 
	The wave characteristics (\figurename{} \ref{fig:PropMK}) all follow similar trends up to $x=7$ m.
	The EVM-$\gamma$ simulation presents the highest wave heights and the HJ-$\sigma$ simulation shows the lowest wave heights.
	After $x=8.4$ m, the skewness and kurtosis in the simulations increase sharply, which may be due to the proximity of the absorption zone (\figurename{} \ref{fig:PropMK}).
	
	When comparing the numerical results with the probe measurements, the time series show some differences in wave phase (\figurename{} \ref{fig:TSMK}), which results in a decrease in the correlations until it reaches $35\%$ for EVM-$\gamma$ and $38\%$ for HJ-$\sigma$ (\figurename{} \ref{fig:PropMK}).
	Nonetheless, the wave asymmetry and skewness are in good agreement despite being slightly underestimated for the simulations, and the NRMSD which remains under $0.20$ (\figurename{} \ref{fig:PropMK}). 
	Between $x=6$ and $8$ m, the simulations underestimate the value of the kurtosis.
	The wave spectra are in good agreement between the simulations and experiments.
	However, the low frequency peak (at $f/f_p \approx 0.12$) is overestimated at some of the probe locations (\figurename{} \ref{fig:FFTMK}).
	
	\clearpage
	
	\subsection{Irregular waves breaking on a submerged slope} \label{subsec:AHA}
		
\begin{figure}[h!]
		\centering
		\includegraphics[width=\linewidth]{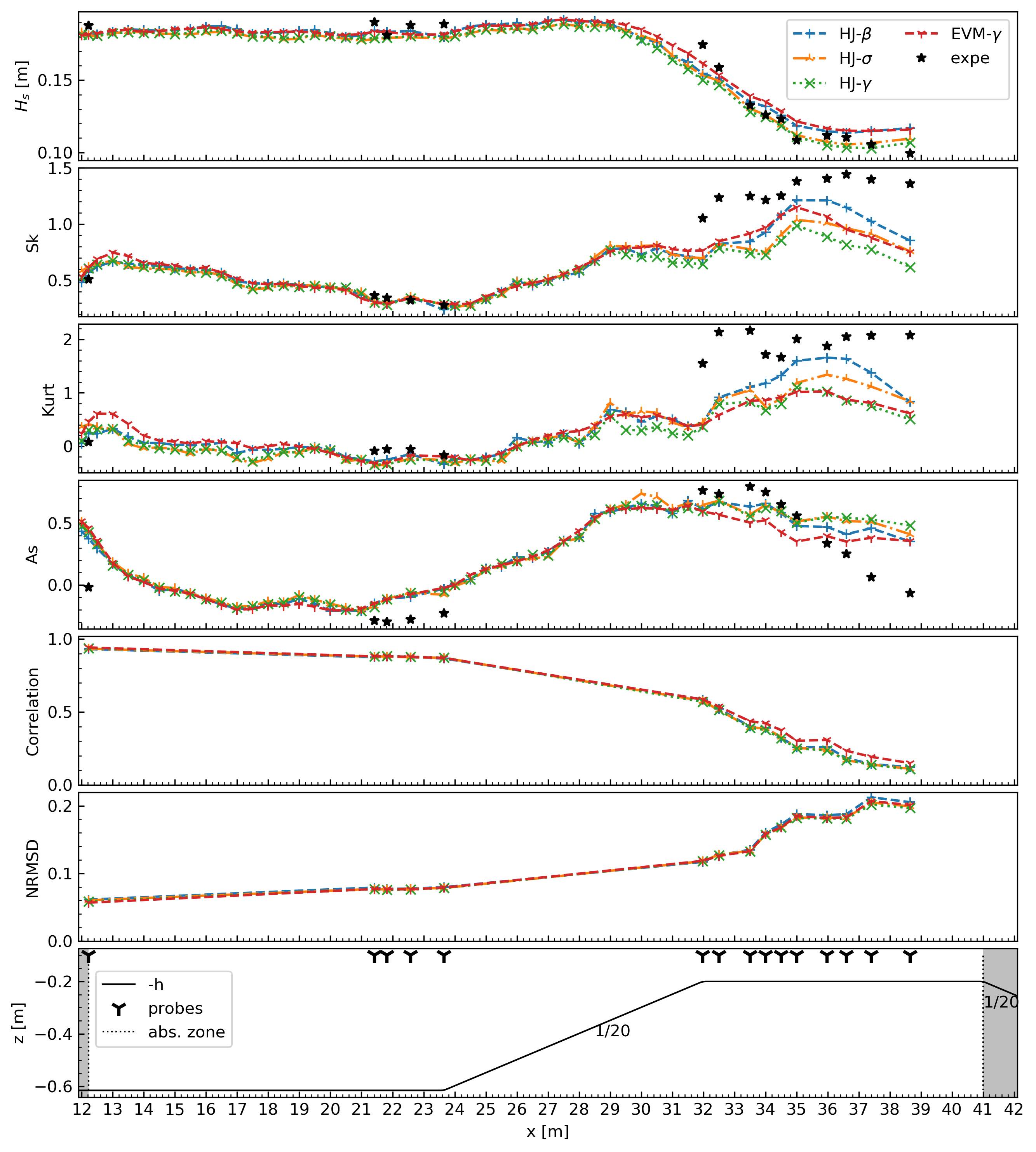}
		\caption{Comparison of the wave properties between the experimental data \citep{adytia_numerical_2018} and numerical simulation results for irregular waves over a submerged slope ($H_s = 0.2$ m, $T_p = 2.5$ s)}
		\label{fig:PropAHA}
	\end{figure}

	The last case focuses on laboratory experiments \citep{husrin_experimental_2012,adytia_numerical_2018} of irregular wave breaking over a submerged slope.
	It was conducted in a channel with a $0.615$ m water depth, reducing to $0.2$ m after a $1$:$20$ slope between $x=23.65$ m and $x=31.98$ m (shown in the lower panel of \figurename{}~\ref{fig:PropAHA}).
	The experiments provide measurements for waves with a peak frequency $f_p=0.4$ Hz and a significant wave height $H_s=0.2$ m.
	Under these wave conditions, the simulations using the HJf dissipation method are not shown here because they became unstable despite trying to progressively apply the dissipation with a ramp coefficient.
	The breaking parameters for the simulations are presented in \tablename{} \ref{tab:AHAref}.
	
	\begin{table}[h!]
		\caption{Wave breaking parameters for the case of irregular waves propagating over a submerged slope \citep{adytia_numerical_2018}}\label{tab:AHAref}
		\centering
		\begin{tabular}{ r|cccc}
    \hline
	Case name  & HJ-$\beta$ & HJ-$\gamma$ & HJ-$\sigma$ & EVM-$\gamma$ \\
	Strength coef. & $\mu=1.1$     & $\mu=1.2$         & $\mu=1.2$    & $\delta=1.$    \\
	Initiation  & $\beta_{i}=30^\circ$ & $\gamma_i=0.5$  & $\sigma_i=0.83$   &   $\gamma_i=0.5$  \\
	Cessation   & $\beta_{t}=10^\circ$ & $\gamma_t=0.15$ & $\beta_{t}=10^\circ$ &  $\gamma_t=0.15$ \\
	\hline
\end{tabular}
	\end{table}

	For this case, the wave skewness, kurtosis and asymmetry increased as the waves propagate upslope.
	The breaking dissipation is activated on average around $x=27$ m, which corresponds to the location of the maximum value reached by $H_s$ (\figurename{} \ref{fig:PropAHA}).	
	
\begin{figure}[h!]
		\centering
		\includegraphics[width=\linewidth]{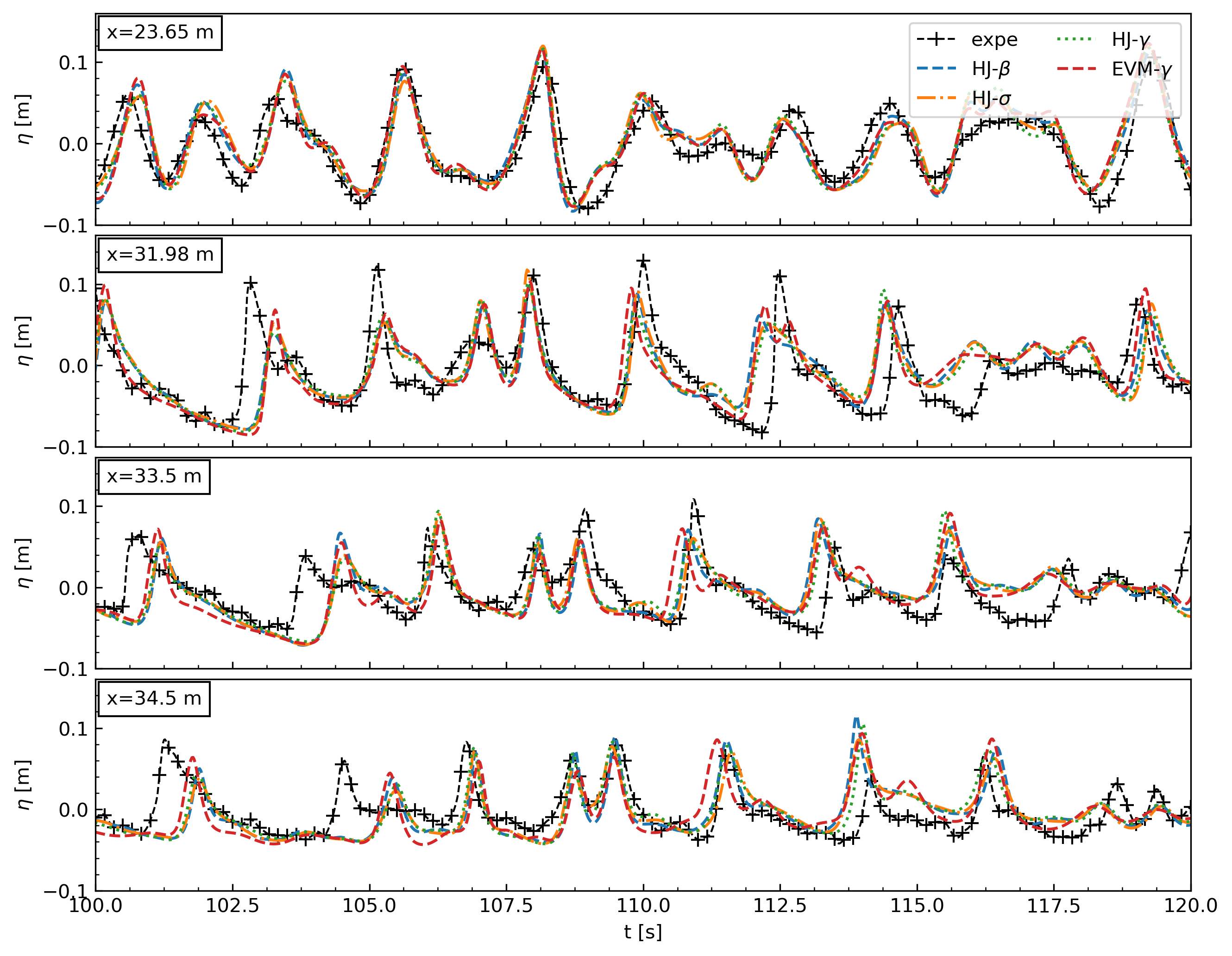}
		\caption{Time series of the free surface elevation for irregular waves breaking over a submerged slope ($H_s = 0.2$ m, $T_p = 2.5$ s). Comparison between the experiments \citep{adytia_numerical_2018} and numerical simulations using different combinations of breaking criteria and dissipation methods}
		\label{fig:TSAHA}
	\end{figure}

\begin{figure}[h!]
		\centering
		\includegraphics[width=\linewidth]{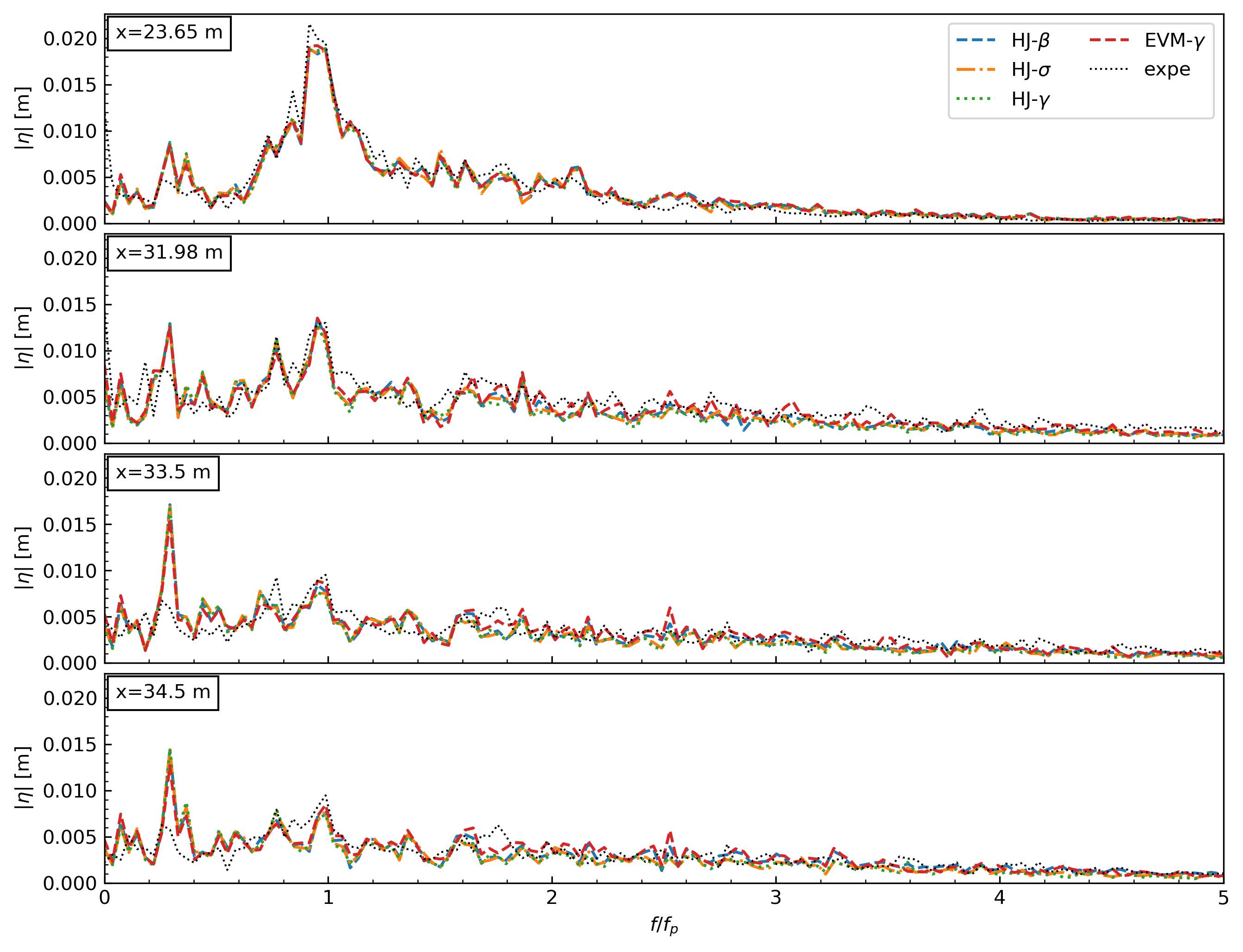}
		\caption{Comparison of the wave amplitude spectra between the experimental data \citep{adytia_numerical_2018} and numerical simulation results for irregular waves over a submerged slope ($H_s = 0.2$ m, $T_p = 2.5$ s)}
		\label{fig:FFTAHA}
	\end{figure}

	The numerical results differ more between themselves in comparison to the previous test cases.	
	After the upslope, the wave shape starts to differ (\figurename{} \ref{fig:TSAHA} and \ref{fig:PropAHA}).
	The wave heights of simulation HJ-$\beta$ and EVM-$\gamma$ are higher than for the two other simulations.
	The skewness, kurtosis and asymmetry of the simulation results also start to differ after  $x=35.5$ m.
	Around $f/f_p=1.8$, $2.6$ and $3.1$ in \figurename{} \ref{fig:FFTAHA} and for $x>31.98$ m, the simulation EVM-$\gamma$ (red curve) shows slightly higher amplitude values than the other simulations.

	Comparing the simulations to the experiments, the largest differences in shape appear after $x=35$ m (\figurename{} \ref{fig:PropAHA}). 
	The asymmetry in the experiments decreases after $x=34$ m, while it remains nearly constant for the other simulations.
	However, the dissipation methods properly capture the decrease of $H_s$.
	The correlation of the time series between simulation EVM-$\gamma$ and the experiments is $5\%$ better than for simulations with HJ dissipation.
	The NRMSD at the last probe of the domain varies between $0.19$ to $0.21$ for HJ-$\gamma$ and HJ-$\beta$ respectively (\figurename{} \ref{fig:PropAHA}).
	
	\clearpage
	
	\section{Conclusion}
		
	The implementation and validation of wave breaking dissipation methods have been presented for the fully nonlinear potential flow solver whispers3D in the case of two dimensional waves.
	Among the various breaking models presented in the literature, three existing dissipation methods were implemented here: HJ and HJf, based on the analogy between a breaking wave and a hydraulic jump \citep{guignard_modeling_2001, grilli_fully_2019}, and an eddy viscosity-like formulation of the dissipation \citep{kennedy_boussinesq_2000}. 
	Three initiation criteria for wave breaking were also compared.
	Those mechanisms proved to be particularly effective for regular wave breaking.
	The time series of the free surface elevation and the spatial evolution of the wave shape were in good agreement with the measurements. 
	In addition, the decomposition of the waves into higher harmonic components during breaking over the trapezoidal bar and the propagation of the wave train after the bar  were also reproduced well.
	For irregular waves, the examined  breaking mechanisms provided less accurate results than for regular waves.
	The simulation results reproduce the free surface correctly during the nonlinear shoaling process. 
	Slight differences with the experiments appeared during the post-breaking evolution.
	Nonetheless, the wave properties and spectra evolution were well predicted by the breaking methods, with differences mainly observed in the lower frequency range. 
	Some spurious low frequency peaks could be observed in the simulation results due to insufficient attenuation of long waves in the relaxation zones used for wave generation and absorption. 
	This effect could be reduced by considering longer relaxation zones and/or techniques more suited to dampen long waves.
	The last case of breaking waves over a submerged slope proved to be more challenging, and the HJf method was unstable, while the other methods provided acceptable results.
	Considering the three irregular wave experiments, the wave spectra over the range $[0.5f_p,\: 5f_p]$ were simulated well before, inside, and after the breaking zone.
	
	The breaking models were selected for their simplicity in practical implementation and their efficiency.
	The main drawback of the breaking methods is the need to calibrate $3$ parameters (whose optimal values appeared to depend on both the wave conditions and the bathymetry): the breaking onset and termination criteria and the strength of the dissipation.
	After calibration of these parameters, the different combinations of breaking criteria and dissipation mechanisms showed equivalent results.
	Some combinations of parameters slightly improved the agreement with the experiments of some wave properties, but the overall comparison of the simulated and experimental time series (assessed by computing the correlation and NRSMD) showed no method being significantly better than another.
	The onset breaking criterion that required the least modification was the local energy flux velocity method.
	This method was recently verified for wave breaking in deep and intermediate depth water \citep{saket_threshold_2017, barthelemy_unified_2018} for values of  $\sigma_i$ close to $0.84$.
	In the presented simulation results, the threshold for breaking gave the best results for $\sigma_i=0.83$ and $0.84$.
	Therefore, the present numerical results support the extension of this criterion to the case of depth-induced wave breaking of irregular wave trains. 

	Overall, the potential flow solver whispers3D proved capable in reproducing wave breaking for both regular and irregular wave conditions, considering three different types of coastal bathymetries.
	Future work will address developments of the code to model run-up on slopes, and the extension of the model to three-dimensional cases to simulate more realistic wave conditions.

	\section*{Acknowledgements}
	
	The authors thank Prof. S. Beji for providing the laboratory data for waves breaking over the bar, Prof. J. T. Kirby for providing the laboratory data of irregular waves breaking on the beach and Dr. S. Husrin for providing laboratory data of irregular waves breaking on the submerged slope.
	
	This work was carried out in the framework of the DEPHYMAN project. 
	It has received funding from Excellence Initiative of Aix-Marseille University - A*MIDEX, a French "Investissements d’Avenir" programme with reference ANR-11-IDEX-0001-02. 
	It has been carried out in the framework of the Labex MEC (Mécanique et Complexité), with reference ANR-10-LABX-0092.
		
	C. Papoutsellis, M. Benoit and M. Yates also acknowledge support from the DiMe project, which benefits from French Government support managed by the ANR under the program "Investissements d'Avenir" with the reference ANR-10-IEED-0006-14.

	\bibliography{Biblio}
	\addcontentsline{toc}{section}{References}
	
\end{document}